\documentclass[pra,twocolumn,showpacs,preprintnumbers,tightenlines,epsfig]{revtex4}
\usepackage{graphicx}
\usepackage{dcolumn}
\usepackage{bm}
\usepackage{amsmath}
\usepackage{amsfonts}
\usepackage{amssymb}

\begin{document}

\title{First-principles quantum simulations of dissociation of molecular
condensates:\\
Atom correlations in momentum space}
\author{C.~M. Savage$^{1,2}$}
\email{craig.savage@anu.edu.au}
\author{P.~E. Schwenn$^{2}$}
\author{K.~V. Kheruntsyan$^{2}$}
\affiliation{$^{1}$ARC Centre of Excellence for Quantum-Atom Optics, Department of
Physics, Australian National University, Canberra ACT 0200, Australia}
\affiliation{$^{2}$ARC Centre of Excellence for Quantum-Atom Optics, School of Physical
Sciences, University of Queensland, Brisbane, QLD 4072, Australia}
\date{\today {}}

\begin{abstract}
We investigate the quantum many-body dynamics of dissociation of a
Bose-Einstein condensate of molecular dimers into pairs of constituent
bosonic atoms and analyze the resulting atom-atom correlations. The quantum
fields of both the molecules and atoms are simulated from first principles
in three dimensions using the positive-$P$ representation method. This
allows us to provide an exact treatment of the molecular field depletion and 
$s$-wave scattering interactions between the particles, as well as to extend
the analysis to nonuniform systems. In the simplest uniform case, we find
that the major source of atom-atom decorrelation is atom-atom recombination
which produces molecules outside the initially occupied condensate mode. The
unwanted\ molecules are formed from dissociated atom pairs with nonopposite
momenta. The net effect of this process -- which becomes increasingly
significant for dissociation durations corresponding to more than about 40\%
conversion -- is to reduce the atom-atom correlations. In
addition, for nonuniform systems we find that mode mixing
due to inhomogeneity can result in further degradation of the correlation
signal. We characterize the correlation strength via the degree of squeezing
of particle number-difference fluctuations in a certain momentum-space
volume and show that the correlation strength can be increased if the
signals are binned into larger counting volumes.
\end{abstract}

\pacs{03.75.Nt,03.65.Ud, 03.75.Gg}
\maketitle


\section{Introduction}
\label{sec: Introduction}

Dissociation of a diatomic molecule produces two quantum mechanically
entangled atoms with equal and opposite momenta in the molecule's rest
frame. These atoms have Einstein-Podolsky-Rosen (EPR) type correlations in
position and momentum \cite{EPR}, and hence are of fundamental interest \cite%
{Kurizki,EPR-Bell-comment}. Experimental advances in coherent manipulation
of quantum gases are now enabling the production of ultracold molecular
gases \cite{Molecules-bosonic-atoms} and even molecular Bose-Einstein
condensates (BECs) \cite{Molecules-fermionic-atoms} from atomic condensates
and degenerate Fermi gases. These molecules can in turn be dissociated \cite%
{Durr,Dissociation-exp-Ketterle,Greiner} into strongly correlated ensembles
of bosonic \cite{Moelmer2001,twinbeams,Yurovsky} or fermionic \cite%
{Jack-Pu,Fermidiss} atoms, thus extending possible fundamental tests of
quantum mechanics into macroscopic regimes \cite{KPRL}. This has close
analogies with experiments in quantum optics using continuous-variable
quadrature correlations generated via parametric down-conversion \cite%
{Ou,eprMDR}.

Apart from the quantum-atom optical aspect of these and related studies
(see, e.g., \cite%
{Phillips-4WM, Roberts,Ketterle-4WM,Meystre-spin-EPR,Duan-spin-EPR,Soerensen-Duan-Zoller,Yurovsky-4WM,Hope,Search-Meystre,Chu-4WM-2005,Ketterle-4WM-2006,Olsen-Davis,Plata-2005,Meystre-diss,Zhao-Astrakharchik}%
), molecular dissociation can serve as a probe of two-body interactions,
including collisional resonances and spectroscopic properties of Feshbach
resonance molecules \cite%
{Wieman-Julienne-dissociation,Rempe-Kokkelmans-dissociation,Braaten-2006,Hanna-2006}%
.

An important recent development in the study of ultracold quantum gases is
the direct measurement of atom-atom correlation functions and quantum
statistics via the analysis of the noise in absorption images \cite%
{Greiner,Bloch,Altman-Lukin,Rzazewski}, atom counting using microchannel
plate detectors \cite{Yasuda-Shimizu,Aspect}, and fluorescence imaging
combined with high-finesse optical cavities \cite{Raizen,Esslinger}. These
experiments have close parallels with the pioneering photon correlation
measurements of Hanbury Brown and Twiss (HBT) \cite{HBT} and initiated
further theoretical studies of HBT interferometry as a sensitive tool for
probing quantum many-body states of ultracold gases \cite%
{Duan-full,Carusotto,Norrie-Ballagh-Gardiner,Rey-Clark,Altman-fermions}. All
these correlation measurements are ultimately related to the quantum-atom
optics counterpart of Glauber's second-order, or density-density,
correlation function \cite{Glauber}. While giving access to the simplest
higher-order correlations, these techniques may in the future lead to more
sophisticated correlation measurements, such as between matter-wave
quadratures required to demonstrate the version of the EPR paradox proposed
in Ref. \cite{KPRL}.

In correlation measurements via the noise in absorption images of
ballistically expanded clouds \cite{Greiner,Bloch}, the extracted
correlation signal reflects the momentum correlations before expansion \cite%
{Altman-Lukin}. This is particularly suitable for studies of molecular
dissociation, since the correlations arise from momentum conservation and
are therefore most fundamentally between atoms with equal and opposite
momenta, $\hbar \mathbf{k}$ and $-\hbar \mathbf{k}$.

In this paper, we study the quantum dynamics of dissociation of a molecular
BEC and analyze the atom-atom correlations in momentum space. Spatial
correlations after expansion will be studied in a subsequent paper \cite%
{SavageKheruntsyanSpatial}. Our analysis is based on first-principles
quantum simulations of the coupled atomic-molecular system in three
dimensions (3D) using the positive $P$-representation method \cite%
{DrummondGardiner,DrummondCarter,Steel,DrummondCorney,DeuarDrummond}. This
allows us to go beyond the analytical results known for the case of a
spatially uniform condensate and undepleted, classical molecular field \cite%
{twinbeams,Fermidiss}, as well as beyond the pairing mean-field method used
in Ref. \cite{Jack-Pu}. In previous related work our simulations were limited to one spatial dimension and to the calculation of the number difference squeezing between all atoms with positive momenta  and all with negative momenta \cite{twinbeams}. In this paper we simulate three spatial dimensions and calculate squeezing localized in momentum space.

In the present treatment, we are able to extend these results and analyze
the molecular depletion without approximations. In addition, the positive-$P$
method allows us to treat nonuniform condensates, as well as $s$-wave
scattering interactions between the species. The main limitation of the
method is that the positive-$P$ simulations eventually fail as the
simulation time increases, particularly when two-body scattering is included 
\cite{Gilchrist}. For this reason, the simulations involving $s$-wave
interactions are performed with slightly reduced values of the scattering
lengths, or else are limited to short dissociation times. Nevertheless, the
results allow us to gain quantitative insights into the role of these
interactions in atom-atom correlations, which have not been analyzed before.

More importantly, we find that mode mixing due to inhomogeneity of the
condensates can potentially have a quite disruptive effect on the atom-atom
correlations. We give quantitative estimates of this effect. Similarly, we
quantify the disruptive role of atom-atom recombination which becomes
increasingly important past the initial spontaneous dissociation regime.

We emphasize that for the correct interpretation of the results on
correlations, especially when comparing the uniform and nonuniform results,
it is important to operate with observables that have a well-defined
operational meaning. We find that the correlation strength is most
conveniently quantified via the normalized number-difference fluctuations.
This is defined with respect to the number of atoms in a certain
momentum-space volume around pairs of the carrier momenta of interest. This
measure of the correlation strength can be defined with respect to different
counting volumes, which corresponds to the procedure of binning often
employed in experiments. We find that the strength of the correlation signal
increases for larger bin sizes, which is in agreement with the results of
analogous experimental measurements of Ref. \cite{Greiner} performed in
position space.

This paper is organized as follows. Section \ref{sec: The model} describes the system we study including relevant approximations. The positive-P simulation method and the full stochastic equations we simulate are presented. Section \ref{analytic} reviews the idealized analytic model presented by Kherutsyan \cite{Fermidiss}. This is the reference against which we can determine the effect of the various realistic processes included in our simulations. Section \ref{numerics} is the heart of the paper and reports numerical simulations that quantify the degradation of quantum correlations in the presence of: molecular depletion (Sec. \ref{depletion}), scattering (Sec. \ref{two-body-terms}), and spatial nonuniformity (Sec. \ref{nonuniform}). The effect of binning data is briefly discussed in Sec. \ref{sec: binning}. The paper ends with a summary, followed by appendices on Fourier transforming the fields.

\section{The model}
\label{sec: The model}

The quantum field theory effective Hamiltonian describing the system we
shall study is given by \cite{twinbeams,PDKKHH-1998}: 
\begin{align}
\hat{H}& =\int d\mathbf{x}\left\{ \sum\nolimits_{i=0,1}\left( \frac{\hbar
^{2}}{2m_{i}}|\mathbf{\nabla }\hat{\Psi}_{i}|^{2}+\hbar V_{i}\hat{\Psi}%
_{i}^{\dagger }\hat{\Psi}_{i}^{{}}\right) \right.  \label{eq:Ham} \\
& \left. -i\frac{\hbar \chi }{2}\left[ \hat{\Psi}_{0}^{\dagger }\hat{\Psi}%
_{1}^{2}-\hat{\Psi}_{1}^{\dagger \,2}\hat{\Psi}_{0}\right]
+\sum\nolimits_{i,j}\frac{\hbar U_{ij}}{2}\hat{\Psi}_{i}^{\dag }\hat{\Psi}%
_{j}^{\dag }\hat{\Psi}_{j}\hat{\Psi}_{i}\right\} .  \notag
\end{align}%
The molecular and atomic fields are, respectively, described by the bosonic
operators $\hat{\Psi}_{0}(\mathbf{x},t)$ and $\hat{\Psi}_{1}(\mathbf{x},t)$
satisfying the following equal-time commutation relation: $[\hat{\Psi}_{i}(%
\mathbf{x},t),\hat{\Psi}_{j}^{\dag }(\mathbf{x}^{\prime },t)]=\delta
_{ij}\delta (\mathbf{x}-\mathbf{x}^{\prime })$ ($i,j=0,1$). The first term
in the Hamiltonian (\ref{eq:Ham}) describes the kinetic energy, where $m_{1}$
and $m_{0}=2m_{1}$ are the atomic and molecular masses. The trapping
potentials, including internal energies $E_{i}$, are given by $V_{i}(\mathbf{%
x})$. This determines the detuning $2\Delta $ which corresponds to the
overall energy mismatch $2\hbar \Delta =\hbar \lbrack
2V_{1}(0)-V_{0}(0)]=2E_{1}-E_{0}$ between the free two-atom state at the
dissociation threshold and the bound molecular state.

The coupling term $\chi $ is responsible for coherent conversion of
molecules into atom pairs, e.g., via Raman transitions or a Feshbach
resonance (see, for example, Refs. \cite%
{PDKKHH-1998,Superchemistry,Feshbach-KKPD,Timmermans,JJ-1999,Holland} and 
Ref. \cite{Stoof-review} for a recent review), while $U_{ij}$ are the strengths
of the two-body $s$-wave interactions describing atom-atom, atom-molecule,
and molecule-molecule scattering. The diagonal terms are given by $%
U_{ii}=4\pi \hbar a_{ii}/m_{i}$, where $a_{11}$ and $a_{00}$ are the
atom-atom and molecule-molecule $s$-wave scattering lengths. The
off-diagonal terms $U_{01}=U_{10}$ are given by $U_{01}=2\pi \hbar
a_{01}/\mu _{01}=3\pi \hbar a_{10}/m_{1}$, where $a_{01}$ is the
atom-molecule $s$-wave scattering length and $\mu
_{01}=m_{0}m_{1}/(m_{0}+m_{1})=2m_{1}/3$ is the reduced mass.

The Hamiltonian (\ref{eq:Ham}) with delta-function interactions implicitly
assumes a momentum cutoff $k_{\max }\lesssim 1/a_{ij}$ \cite{Abrikosov}, and
we associate the coupling terms with the observed scattering lengths. In the
numerical simulations on a finite computational lattice, where there is
always a maximum cutoff, this description is valid as long as the lattice
spacing in coordinate space is larger than $a_{ij}$. Otherwise, a more
careful renormalization procedure \cite{Holland} is needed.

Starting from a stable ($E_{0}<2E_{1}$) molecular BEC, dissociation into the
constituent atoms may be achieved by a rapid Feshbach sweep into the atomic
side of the resonance or by a coherent Raman transition \cite%
{twinbeams,Durr,Greiner}. If the transition has a negative detuning $2\hbar
\Delta <0$ after the sweep ($E_{0}>2E_{1}$), the molecules become unstable
against (spontaneous) dissociation into free atom pairs. The energy level
configuration after the sweep is the initial condition for our simulations.
We assume that the molecular BEC is initially in a coherent state, whereas
the atoms are in the vacuum state. In addition, we assume that once the
dissociation coupling $\chi $ is switched on, the trapping potentials are
simultaneously switched off, so that the evolution of the atomic and
molecular fields takes place in free space.

For molecules at rest, the excess of potential energy $2\hbar |\Delta |$ is
converted into kinetic energy, $2\hbar |\Delta |\rightarrow 2\hbar
^{2}k^{2}/(2m_{1})$, of atom pairs with equal but opposite momenta around $%
\pm \mathbf{k}_{0}$, where $k_{0}=|\mathbf{k}_{0}|=\sqrt{2m_{1}|\Delta
|/\hbar }$. This is the physical origin of the expected correlations between
the atoms, which we will study below in greater detail and in the context
of many-body field theory.

To investigate the quantum dynamics we use stochastic differential equations
corresponding to the positive-$P$ representation of the density matrix \cite%
{DrummondGardiner,DrummondCarter,Steel,DrummondCorney,DeuarDrummond}. The
essence of this method is the mapping of the operator equations of motion
into $c$ number stochastic differential equations that can be solved
numerically. This requires two complex stochastic fields $\Psi
_{i}(\mathbf{x},t)$ and $\Phi _{i}(\mathbf{x},t)$ [$i=0,1$, $\Phi _{i}^{\ast
}(\mathbf{x},t)\neq \Psi _{i}(\mathbf{x},t)$], corresponding to the
operators $\hat{\Psi}_{i}(\mathbf{x},t)$ and $\hat{\Psi}_{i}^{\dag }(\mathbf{%
x},t)$, respectively. Ensemble averages of the stochastic fields over a
large number of trajectories $\langle \ldots \rangle _{\mathrm{st}}$
correspond to quantum mechanical ensemble averages of normally ordered
operator moments. For example%
\begin{equation}
\langle \lbrack \hat{\Psi}_{i}^{\dagger }(\mathbf{x},t)]^{k}[\hat{\Psi}_{j}(%
\mathbf{x}^{\prime },t)]^{n}\rangle =\langle \lbrack \Phi _{i}(\mathbf{x}%
,t)]^{k}[\Psi _{j}(\mathbf{x}^{\prime },t)]^{n}\rangle _{\mathrm{st}}.
\end{equation}

The stochastic differential equations governing the quantum dynamical
evolution under the Hamiltonian (\ref{eq:Ham}) are, in a rotating frame:%
\begin{widetext}%
\begin{align}
\frac{\partial\Psi_{1}}{\partial t} &  =\frac{i\hbar}{2m_{1}}\mathbf{\nabla
}^{2}\Psi_{1}-i\left(  \Delta+\sum\nolimits_{i}U_{1i}\Phi_{i}\Psi_{i}\right)
\Psi_{1}+\chi\Psi_{0}\Phi_{1}+\sqrt{\chi\Psi_{0}}\;\zeta_{1}+\sqrt
{-iU_{01}\Psi_{1}\Psi_{0}/2}\;(\zeta_{2}+i\zeta_{3})+\sqrt{-iU_{11}\Psi
_{1}^{2}}\;\zeta_{4},\nonumber\\
\frac{\partial\Phi_{1}}{\partial t} &  =-\frac{i\hbar}{2m_{1}}\mathbf{\nabla
}^{2}\Phi_{1}+i\left(  \Delta+\sum\nolimits_{i}U_{1i}\Phi_{i}\Psi_{i}\right)
\Phi_{1}+\chi\Phi_{0}\Psi_{1}+\sqrt{\chi\Phi_{0}}\;\zeta_{5}+\sqrt{iU_{01}%
\Phi_{1}\Phi_{0}/2}\;(\zeta_{6}+i\zeta_{7})+\sqrt{iU_{11}\Phi_{1}^{2}}%
\;\zeta_{8},\nonumber\\
\frac{\partial\Psi_{0}}{\partial t} &  =\frac{i\hbar}{2m_{0}}\mathbf{\nabla
}^{2}\Psi_{0}-i\sum\nolimits_{i}U_{0i}\Phi_{i}\Psi_{i}\Psi_{0}-\frac{\chi}%
{2}\Psi_{1}^{2}+\sqrt{-iU_{01}\Psi_{1}\Psi_{0}/2}\;(\zeta_{2}-i\zeta
_{3})+\sqrt{-iU_{00}\Psi_{0}^{2}}\;\zeta_{9},\nonumber\\
\frac{\partial\Phi_{0}}{\partial\tau} &  =-\frac{i\hbar}{2m_{0}}%
\mathbf{\nabla}^{2}\Phi_{0}+i\sum\nolimits_{i}U_{0i}\Phi_{i}\Psi_{i}\Phi
_{0}-\frac{\chi}{2}\Phi_{1}^{2}+\sqrt{iU_{01}\Phi_{1}\Phi_{0}/2}\;(\zeta
_{6}-i\zeta_{7})+\sqrt{iU_{00}\Phi_{0}^{2}}\;\zeta_{10},\label{Positive-P-eqs}%
\end{align}
\end{widetext}Here, the noise terms $\zeta _{j}$ ($j=1,...,10$) are real,
independent Gaussian noises with $\langle \zeta _{j}(\mathbf{x},t)\rangle _{%
\mathrm{st}}=0$ and the following nonzero correlations $\langle \zeta _{j}(%
\mathbf{x},t)\zeta _{k}(\mathbf{x}^{\prime },t^{\prime })\rangle _{\mathrm{st%
}}=\delta _{jk}\delta (\mathbf{x}-\mathbf{x}^{\prime })\delta (t-t^{\prime
}) $.

We note that the Hamiltonian (\ref{eq:Ham}) conserves the total number of
atomic particles%
\begin{equation}
\hat{N} = \hat{N}_{1}(t)+2 \hat{N}_{0}(t),
\end{equation}%
whether they are counted as bound pairs in the molecular state or as free
atoms, $\hat{N}_{i}(t)=\int d\mathbf{x}  \hat{\Psi}_{i}^{\dagger }(\mathbf{x}%
,t)\hat{\Psi}_{i}^{{}}(\mathbf{x},t)  $ ($i=0,1$). Since we are
interested in dissociation of a pure molecular BEC, with no atoms present
initially, then $\hat{N}$ is a known constant operator given by $\hat{N} =2 \hat{N}_{0}(0)$, where $%
\hat{N}_{0}(0)$ is the total initial number of molecules. Expectation values of these operators will be denoted by removing the hats, e.g. $N_i = \langle \hat{N}_i \rangle$.

\section{Analytic solutions in the uniform system}
\label{analytic}

The simplest treatment of dissociation resulting in analytic solutions is
achieved under an undepleted molecular field approximation in a uniform
system \cite{twinbeams,Fermidiss}. The approximation is valid for short
dissociation times during which the total number of atoms produced is only a
small fraction ($\lesssim 10$\%) of the initial number of molecules.

In this treatment we consider a uniform molecular BEC in a coherent state
with a constant amplitude $\Psi _{0}=\sqrt{\rho _{0}}$ (where $\rho _{0}$ is
the uniform density), which we assume is real without loss of generality.
The condensate is contained within a cubic box of side $L$ (volume $=L^{3}$)
with periodic boundary conditions, extending from $-L/2$ to $L/2$ in each
dimension. The dissociation coupling $\chi $ is turned on suddenly, and
subsequently assumed to be constant.

The Heisenberg equations for the atomic field operators are then 
\begin{align}
\frac{\partial \hat{\Psi}_{1}(\mathbf{x},t)}{\partial t}& =i \frac{\hbar}{2m_1} \mathbf{\nabla }%
^{2}\hat{\Psi}_{1}-i\Delta _{\mathrm{eff}}\hat{\Psi}_{1}+g\hat{\Psi}%
_{1}^{\dag },  \notag \\
\frac{\partial \hat{\Psi}_{1}^{\dagger }(\mathbf{x},t)}{\partial t}& =-i \frac{\hbar}{2m_1} %
\mathbf{\nabla }^{2}\hat{\Psi}_{1}^{\dagger }+i\Delta _{\mathrm{eff}}\hat{%
\Psi}_{1}^{\dagger }+g\hat{\Psi}_{1},  \label{Heisenberg-eqs}
\end{align}%
where $g\equiv \chi \sqrt{\rho _{0}}$ is the coupling and $\Delta _{\mathrm{%
eff}}\equiv \Delta +(U_{01}-U_{00}/2)\rho _{0}$ is the effective detuning
that takes into account the initial mean-field energy shifts due to
atom-molecule and molecule-molecule interactions. We note that in our
simulations we will be interested in cases with large bare\ detuning $%
|\Delta |\gg |U_{01}-U_{00}/2|\rho _{0}$ so that the mean-field energy
shifts are negligible.

Introducing creation and annihilation operators $\hat{a}_{\mathbf{k}}(t)$
and $\hat{a}_{\mathbf{k}}^{\dagger }(t)$ for the plane-wave atomic modes
with momentum $\mathbf{k}$ and commutation relations $[\hat{a}_{\mathbf{k}%
}(t),\hat{a}_{\mathbf{k}^{\prime }}^{\dagger }(t)]=\delta _{\mathbf{k},%
\mathbf{k}^{\prime }}$ (see Appendix A), the Heisenberg equations (\ref%
{Heisenberg-eqs}) transform into 
\begin{align}
\frac{d\hat{a}_{\mathbf{k}}(t)}{dt}& =-i\left( \frac{\hbar \mathbf{k}^{2}}{%
2m_{1}}+\Delta _{\mathrm{eff}}\right) \hat{a}_{\mathbf{k}}(t)+g\hat{a}_{-%
\mathbf{k}}^{\dag },  \notag \\
\frac{d\hat{a}_{-\mathbf{k}}^{\dag }}{dt}& =i\left( \frac{\hbar \mathbf{k}%
^{2}}{2m_{1}}+\Delta _{\mathrm{eff}}\right) \hat{a}_{-\mathbf{k}}^{\dag }+g%
\hat{a}_{\mathbf{k}}(t).  \label{Heisenberg-eqs-k-space}
\end{align}

These have the following solution 
\begin{align}
\hat{a}_{\mathbf{k}}(t)& =A_{k}(t)\hat{a}_{\mathbf{k}}(0)+B_{k}(t)\hat{a}_{-%
\mathbf{k}}^{\dag }(0),  \notag \\
\hat{a}_{-\mathbf{k}}^{\dag }(t)& =B_{k}(t)\hat{a}_{\mathbf{k}%
}(0)+A_{k}^{\ast }(t)\hat{a}_{-\mathbf{k}}^{\dag }(0),
\label{Heisenberg-eqs-k-space-soln}
\end{align}%
where the coefficients are 
\begin{align}
A_{k}(t)& =\cosh (g_{k}t)-i\Delta _{k}\sinh (g_{k}t)/g_{k},  \notag \\
B_{k}(t)& =g\;\sinh (g_{k}t)/g_{k},  \label{coefficients}
\end{align}%
and we have introduced $\Delta _{k}\equiv \hbar \mathbf{k}%
^{2}/(2m_{1})+\Delta _{\mathrm{eff}}$ and $g_{k}\equiv \lbrack g^{2}-\Delta
_{k}^{2}]^{1/2}$, where $k=|\mathbf{k}|$. The coefficients $A_{k}(t)$ and $%
B_{k}(t)$ satisfy $|A_{k}(t)|^{2}-B_{k}^{2}(t)=1$.

Given the initial vacuum state of the atomic field we can calculate any
expectation values of the field operators. In particular, we find that the
only nonzero second-order moments are the particle number per mode and the
pairing field per mode: 
\begin{align}
n_{\mathbf{k}}(t)& =\langle \hat{n}_{\mathbf{k}}(t)\rangle
=B_{k}^{2}(t)=(g/g_{k})^{2}\;\sinh ^{2}(g_{k}t),  \label{n-moment} \\
m_{\mathbf{k}}(t)& =\langle \hat{m}_{\mathbf{k}}(t)\rangle =A_{k}(t)B_{k}(t),
\label{m-moment}
\end{align}%
where $\hat{n}_{\mathbf{k}}(t)=\hat{a}_{\mathbf{k}}^{\dagger }(t)\hat{a}_{%
\mathbf{k}}(t)$ and $\hat{m}_{\mathbf{k}}(t)=\hat{a}_{\mathbf{k}}(t)\hat{a}%
_{-\mathbf{k}}(t)$ are the respective operators. All other second-order
moments are equal to zero. Higher-order moments factorize according to
Wick's theorem and can be expressed in terms of the above second-order
moments. From $|A_{k}(t)|^{2}-B_{k}^{2}(t)=1$, we find in addition that $n_{%
\mathbf{k}}(t)$ and $m_{\mathbf{k}}(t)$ are related by 
\begin{equation}
|m_{\mathbf{k}}(t)|^{2}=n_{\mathbf{k}}(t)[1+n_{\mathbf{k}}(t)].
\label{m-n relation}
\end{equation}

Using these solutions, the total number of atoms produced is 
\begin{equation}
N_{1}(t)=\sum\nolimits_{\mathbf{k}}n_{\mathbf{k}}(t)=\sum\nolimits_{\mathbf{k%
}}B_{k}^{2}(t).  \label{total-atom-number}
\end{equation}

The treatment of an infinite system in free space is achieved as usual by
taking the limit of $L\rightarrow \infty $ \ ($\Delta k\rightarrow 0$) and
transforming the mode creation and annihilation operators into continuous
Fourier transforms $\hat{a}(\mathbf{k})$ and $\hat{a}^{\dagger }(\mathbf{k})$
that satisfy the commutation relation $[\hat{a}(\mathbf{k}),\hat{a}^{\dagger
}(\mathbf{k}^{\prime })]=\delta (\mathbf{k}-\mathbf{k}^{\prime })$ (see
appendix A). In this case, the solutions for the normal and anomalous
moments are 
\begin{align}
\langle \hat{a}^{\dagger }(\mathbf{k},t)\hat{a}(\mathbf{k}^{\prime
},t)\rangle & =B_{k}^{2}(t)\delta (\mathbf{k}-\mathbf{k}^{\prime }), \\
\langle \hat{a}(\mathbf{k},t)\hat{a}(\mathbf{k}^{\prime },t)\rangle &
=A_{k}(t)B_{k}(t)\delta (\mathbf{k}+\mathbf{k}^{\prime }).
\end{align}

\subsection{Atom-atom correlations}

\label{correlations}

We are interested in the atomic correlations resulting from momentum
conservation in the molecular dissociation. These may be quantified using a
number of different, but related, density-density correlation functions.
Furthermore, these measures may correlate fields at points in position space
or momentum space. Since the fundamental correlation is between momenta, the
momentum correlation functions are simpler to interpret. However,
experimentally it is easier to measure spatial correlations after
time-of-flight expansion of the cloud. In the present paper we will
concentrate on analyzing atomic correlations in momentum space. 
Spatial correlations after expansion will be studied in a future
work \cite{SavageKheruntsyanSpatial}.

Starting from the atomic vacuum, and assuming no initial momentum spread in
the molecular BEC and no scattering, molecular dissociation produces pairs
of atoms with equal and opposite momenta. In the short time limit this is
well approximated by the analytic solutions of the previous subsection. The
strength of the atomic correlations may be quantified via Glauber's
second-order correlation function \cite{Glauber} 
\begin{equation}
g^{(2)}(\mathbf{k},\mathbf{k^{\prime }},t)=\frac{\langle \hat{a}_{\mathbf{k}%
}^{\dagger }(t)\hat{a}_{\mathbf{k}^{\prime }}^{\dagger }(t)\hat{a}_{\mathbf{k%
}^{\prime }}(t)\hat{a}_{\mathbf{k}}(t)\rangle }{\langle \hat{n}_{\mathbf{k}%
}(t)\rangle \;\langle \hat{n}_{\mathbf{k}^{\prime }}(t)\rangle }.  \label{g2}
\end{equation}%
It is defined in terms of normally ordered operator products and is
normalized so that it is dimensionless. The pair correlation describes the
ratio of the probability of joint detection of pairs of atoms with $\mathbf{k%
}$ and $\mathbf{k}^{\prime }$ to the product of probabilities of independent
atom detection events at $\mathbf{k}$ and $\mathbf{k}^{\prime }$. For
example, $g^{(2)}=1$ for uncorrelated atoms, $g^{(2)}=2$ for thermally
bunched atoms, and $g^{(2)}>2$ indicates super-thermal bunching.

Note that the conditions for an expansion using Wick's theorem are met by
the solutions for the uniform system under the undepleted molecular field
approximation, Eq.~(\ref{Heisenberg-eqs-k-space}). In this case, the pair
correlation function $g^{(2)}(\mathbf{k},\mathbf{k}^{\prime },t)$ can be
factorized and be expressed in terms of products of the second-order normal
and anomalous moments $\langle \hat{a}_{\mathbf{k}}^{\dagger }(t)\hat{a}_{%
\mathbf{k}^{\prime }}(t)\rangle $ and $\langle \hat{a}_{\mathbf{k}}(t)\hat{a}%
_{\mathbf{k}^{\prime }}(t)\rangle $. For example, in the case of $\mathbf{k}%
^{\prime }=-\mathbf{k}$, the pair correlation function is 
\begin{align}
g^{(2)}(\mathbf{k},-\mathbf{k},t)& =1+\frac{|\langle \hat{a}_{\mathbf{k}%
}^{\dagger }(t)\hat{a}_{-\mathbf{k}}(t)\rangle |^{2}}{\langle \hat{n}_{%
\mathbf{k}}(t)\rangle ^{2}}+\frac{|\langle \hat{a}_{\mathbf{k}}(t)\hat{a}_{-%
\mathbf{k}}(t)\rangle |^{2}}{\langle \hat{n}_{\mathbf{k}}(t)\rangle ^{2}} 
\notag \\
& =1+\frac{|m_{\mathbf{k}}(t)|^{2}}{n_{\mathbf{k}}(t)^{2}}=2+\frac{1}{n_{%
\mathbf{k}}(t)},  \label{g2kmk}
\end{align}%
where we have taken into account that $\langle \hat{n}_{\mathbf{k}%
}(t)\rangle =\langle \hat{n}_{-\mathbf{k}}(t)\rangle $, $\langle \hat{a}_{%
\mathbf{k}}^{\dagger }(t)\hat{a}_{-\mathbf{k}}(t)\rangle =0$ and used Eq. (%
\ref{m-n relation}). For other pairs of momenta the correlation function is
obtained in a similar way, with the overall result given by 
\begin{equation}
g^{(2)}(\mathbf{k},\mathbf{k}^{\prime },t)=\left\{ 
\begin{array}{l}
1,\,\;\quad \quad \quad \quad \,\,\mathbf{k}^{\prime }\neq \mathbf{k},%
\mathbf{k}^{\prime }\neq -\mathbf{k}\\ 
2,\,\;\quad \quad \quad \quad \,\,\mathbf{k}^{\prime }=\mathbf{k}\neq 0, \\ 
2+1/n_{\mathbf{k}}(t),\,\,\mathbf{k}^{\prime }=-\mathbf{k},\\ 
3+1/n_{\mathbf{k}}(t),\,\,\mathbf{k}^{\prime }=\mathbf{k}=0%
\end{array}%
\right. .  \label{analytic correlation functions}
\end{equation}

As we see, the correlation function for atom pairs with equal but opposite
momenta is the strongest [except for the special case of $\mathbf{k}=\mathbf{k}%
^{\prime }=0$]. Quantitatively, strong superbunching in $g^{(2)}(\mathbf{k}%
,-\mathbf{k},t)$ is achieved for low mode populations, $n_{\mathbf{k}}(t)\ll
1$. In this regime, the numerator in Eq.~(\ref{g2}) is approximately
proportional to the mode population $n_{\mathbf{k}}(t)$, while the
denominator is the product of populations, and therefore $g^{(2)}(\mathbf{k}%
,-\mathbf{k},t)\approx 1/n_{\mathbf{k}}(t)\gg 1$. As the mode occupancies
grow with time, the pair correlation 
approaches the level of thermal bunching $g^{(2)}(\mathbf{k},-\mathbf{k}%
,t)\rightarrow 2$.

Thus, at high densities Glauber pair correlation becomes a less sensitive
measure of the correlation strength. This is despite the fact that the
correlation between the atoms with equal and opposite momenta is still
maximal at any given density, in this analytically soluble model. There is
perfect squeezing of particle number-difference fluctuations below the
shot-noise level. This is quantified via the normalized variance 
\begin{gather}
V(\mathbf{k},\mathbf{k}^{\prime },t)=\frac{\langle \left[ \Delta \left( \hat{%
n}_{\mathbf{k}}(t)-\hat{n}_{\mathbf{k}^{\prime }}(t)\right) \right]
^{2}\rangle }{\langle \hat{n}_{\mathbf{k}}(t)\rangle +\langle \hat{n}_{%
\mathbf{k}^{\prime }}(t)\rangle }  \notag \\
=1+\frac{\langle :\left[ \Delta \left( \hat{n}_{\mathbf{k}}(t)-\hat{n}_{%
\mathbf{k}^{\prime }}(t)\right) \right] ^{2}:\rangle }{\langle \hat{n}_{%
\mathbf{k}}(t)\rangle +\langle \hat{n}_{\mathbf{k}^{\prime }}(t)\rangle },
\label{variance}
\end{gather}%
where $\Delta \hat{A}=\hat{A}-\langle \hat{A}\rangle $ is the fluctuation
and the colons $::$ indicate normal ordering of the creation operators
before the annihilation operators. This definition uses the conventional normalization with respect to the
shot-noise level resulting from a Poissonian number probability
distribution, such as for a coherent state. This is given by the sum of the
mean occupation numbers, $\langle \hat{n}_{\mathbf{k}}(t)\rangle +\langle 
\hat{n}_{\mathbf{k}^{\prime }}(t)\rangle $. Variance smaller than one, $V(%
\mathbf{k},\mathbf{k}^{\prime },t)<1$, implies reduction of fluctuations
below the shot-noise level and is due to correlation between particle number
fluctuations in the $\mathbf{k}$ and $\mathbf{k}^{\prime }$-modes, while $V(%
\mathbf{k},\mathbf{k}^{\prime },t)=1$ for uncorrelated modes.

For the solutions of Eq.~(\ref{Heisenberg-eqs-k-space}) the
number-difference variance for equal but opposite momenta is given by $V(%
\mathbf{k},-\mathbf{k},t)=0$ at \textit{all} times, i.e., for all occupation
numbers $n_{\mathbf{k}}(t)=n_{-\mathbf{k}}(t)$. This is the ideal case,
implying perfect correlation between fluctuations in $\hat{n}_{\mathbf{k}%
}(t) $ and $\hat{n}_{-\mathbf{k}}(t)$ and corresponding to $100\%$ squeezing
below the shot-noise level.

We note that for two equally occupied modes, the variance $V(\mathbf{k},-%
\mathbf{k},t)$ and the second-order correlation function $g^{(2)}(\mathbf{k}%
,-\mathbf{k},t)$ are related by%
\begin{equation}
V(\mathbf{k},-\mathbf{k},t)=1-n_{\mathbf{k}}(t)[g^{(2)}(\mathbf{k},-\mathbf{k%
},t)-g^{(2)}(\mathbf{k},\mathbf{k},t)],  \label{Vg relation}
\end{equation}
where we have used $g^{(2)}(\mathbf{k},\mathbf{k},t)=g^{(2)}(-\mathbf{k},-%
\mathbf{k},t)$ due to the spherical symmetry of the problem. Therefore,
perfect noise reduction, $V(\mathbf{k},-\mathbf{k},t)=0$, implies that the
maximum degree of correlation is $g_{\max}^{(2)}(\mathbf{k},-\mathbf{k},t)=$ 
$g^{(2)}(\mathbf{k},\mathbf{k},t)+1/n_{\mathbf{k}}(t)$. Using the thermal
level of autocorrelation that occurs in the analytic solutions, Eq.(\ref{analytic correlation functions}), 
$g^{(2)}(\mathbf{k},\mathbf{k},t)=2$, this
gives%
\begin{equation}
g_{\max}^{(2)}(\mathbf{k},-\mathbf{k},t)=2+1/n_{\mathbf{k}}(t),
\label{g2max}
\end{equation}
which coincides exactly with the analytic correlation, Eq. (\ref%
{analytic correlation functions}).

In a similar way, one can show that the maximum (ideal) degree of
correlation Eq.~(\ref{g2max}) corresponds to the maximum anomalous moment $%
m_{\mathbf{k}}(t)$%
\begin{equation}
\max \{ | m_{\mathbf{k}}(t) |^2 \}=n_{\mathbf{k}}(t)[1+n_{\mathbf{k}}(t)],
\label{max-m}
\end{equation}%
which again coincides with the solution, Eq.~(\ref{m-n relation}), of the
present analytically soluble model. This result follows immediately from the
Wick-factorized expression for $g^{(2)}(\mathbf{k},-\mathbf{k},t)$, Eq.~(\ref%
{g2kmk}), 
\begin{equation}
g^{(2)}(\mathbf{k},-\mathbf{k},t)=1+\frac{|m_{\mathbf{k}}(t)|^{2}}{n_{%
\mathbf{k}}(t)^{2}},  \label{Wick-m}
\end{equation}%
assuming $\langle \hat{a}_{\mathbf{k}}^{\dagger }(t)\hat{a}_{-\mathbf{k}%
}(t)\rangle =0$.

More generally, i.e., for nonideal cases, the number-difference variance $V(%
\mathbf{k},-\mathbf{k},t)$ would be larger than zero and can serve as a
sensitive measure of the correlation strength between particle number
fluctuations in different modes. In the nonideal cases, $V(\mathbf{k},-%
\mathbf{k},t)>0$ also implies that the following inequalities hold:%
\begin{eqnarray}
g^{(2)}(\mathbf{k},-\mathbf{k},t) &<&2+1/n_{\mathbf{k}}(t),
\label{g2-nonideal} \\
| m_{\mathbf{k}}(t) |^2 &<&n_{\mathbf{k}}(t)[1+n_{\mathbf{k}}(t)],
\label{m-nonideal}
\end{eqnarray}%
provided that one still has $g^{(2)}(\mathbf{k},\mathbf{k},t)=2$ and $%
\langle \hat{a}_{\mathbf{k}}^{\dagger }(t)\hat{a}_{-\mathbf{k}}(t)\rangle =0$%
.

Quantifying the strength of correlations via the number-difference variance $%
V(\mathbf{k},-\mathbf{k},t)$ is especially useful at large mode occupancies $%
n_{\mathbf{k}}(t)\gg 1$, when the pair correlation 
approaches $g^{(2)}(\mathbf{k},-\mathbf{k},t)\rightarrow 2$
and therefore becomes a less sensitive measure of the correlation strength.

\section{Numerical results and discussion}

\label{numerics}

Our goal in the analysis of atom correlations in molecular dissociation is
to go beyond the undepleted molecular field approximation and to treat the
quantum dynamics for realistic nonuniform systems. Analytic solutions for
these cases are no longer available, and we will resort to numerical
solution of the positive-$P$ equations (\ref{Positive-P-eqs}), which
describe all quantum effects from first principles.

Using the positive-$P$ method, we are able to extend the analytic results of
the previous section in three ways, to include: (a) the depletion of the
molecular condensate; (b) $s$-wave scattering interactions; and (c)
nonuniform condensates.

In the first two cases, we will consider uniform systems, while the
nonuniform case will be simulated within the undepleted molecular field
approximation. The reason for doing this is to give a precise quantitative
description of \textit{each} of these effects and to assess their relative
importance in altering the atom-atom correlations. In addition, the results
obtained in each case can serve as a benchmark for other approximate
numerical methods. An example is the pairing mean-field method of Ref. \cite%
{Jack-Pu} which treats the molecular field depletion at the level of a
coherent state, while the dynamics of the atomic field is treated via the
normal and anomalous densities. The range of validity of this method will be
examined below using the exact results of the present treatment.

The positive-$P$ numerical method is implemented on a uniform spatial
lattice with periodic boundary conditions, and the continuous fields are
therefore represented by discrete mode amplitudes. The momentum space
lattice is reciprocal to the spatial lattice, with $\Delta k=2\pi /L$ being
the lattice spacing and $L$ the length of the spatial domain in each
dimension.

Since the numerical method is able to treat both uniform systems in a finite
box as well as nonuniform systems in free space, the momentum-space field
amplitudes are commonly treated via the lattice-discretized momentum
components $\hat{a}(\mathbf{k},t)$ and $\hat{a}^{\dagger }(\mathbf{k},t)$,
which correspond to the continuous Fourier transforms (see Appendix A) in
the limit $\Delta k\rightarrow 0$ ($L\rightarrow \infty $). (The
correspondences between the operator Fourier components and those of the
stochastic fields are outlined in Appendix B). Accordingly, for any finite
computational lattice one can formally identify a set of momentum modes
described by creation and annihilation operators $\hat{a}_{\mathbf{k}}(t)$
and $\hat{a}_{\mathbf{k}}^{\dagger }(t)$ (with commutators $[\hat{a}_{%
\mathbf{k}}(t),\hat{a}_{\mathbf{k}^{\prime }}^{\dagger }(t)]=\delta _{%
\mathbf{k},\mathbf{k}^{\prime }}$), which are related to their
lattice-discretized continuous counterparts $\hat{a}(\mathbf{k},t)$ and $%
\hat{a}^{\dagger }(\mathbf{k},t)$ via $\hat{a}_{\mathbf{k}}(t)=\hat{a}(%
\mathbf{k},t)\left(\Delta k\right)^{3/2}$ and $\hat{a}_{\mathbf{k}}^{\dagger
}(t)=\hat{a}^{\dagger }(\mathbf{k},t)\left( \Delta k\right)^{3/2}$.

Similarly, the (dimensionless) mode populations $n_{\mathbf{k}}(t)$ and the
pairing fields per mode $m_{\mathbf{k}}(t)$ are related to the\ normal and
anomalous densities $n(\mathbf{k},t)=\langle \hat{n}(\mathbf{k},t)\rangle
=\langle \hat{a}^{\dagger }(\mathbf{k},t)\hat{a}(\mathbf{k},t)\rangle $ and $%
m(\mathbf{k},t)=\langle \hat{m}(\mathbf{k},t)\rangle =\langle \hat{a}(%
\mathbf{k},t)\hat{a}(-\mathbf{k},t)\rangle $ [having units of m$^{3}$] via%
\begin{align}
n_{\mathbf{k}}(t)& =n(\mathbf{k},t)\left( \Delta k\right) ^{3},
\label{n-cont-discr} \\
m_{\mathbf{k}}(t)& =m(\mathbf{k},t)\left( \Delta k\right) ^{3}.
\label{m-cont-discr}
\end{align}

In the continuous limit, these should be understood as corresponding to the
expectation values of the respective operators defined as integrals over the
momentum-space lattice volume element $v(\mathbf{k})=\left( \Delta k\right)
^{3}$ around $\mathbf{k}$, 
\begin{align}
\hat{n}_{\mathbf{k}}(t)& =\int_{v(\mathbf{k})}\hat{n}(\mathbf{k}^{\prime
},t)d\mathbf{k}^{\prime }\;\mathbf{\simeq }\;\hat{n}(\mathbf{k},t)\left(
\Delta k\right) ^{3},  \label{n-cont} \\
\hat{m}_{\mathbf{k}}(t)& =\int_{v(\mathbf{k})}\hat{m}(\mathbf{k}^{\prime
},t)d\mathbf{k}^{\prime }\;\mathbf{\simeq }\;\hat{m}(\mathbf{k},t)\left(
\Delta k\right) ^{3},  \label{m-cont}
\end{align}%
where $\Delta k$ is small enough so that $\hat{n}(\mathbf{k}^{\prime },t)$
and $\hat{m}_{\mathbf{k}}(\mathbf{k}^{\prime },t)$ under the integrals do
not vary much within the integration volume and can be replaced by $\hat{n}(%
\mathbf{k},t)$ and $\hat{m}(\mathbf{k},t)$.

The total number of atoms is 
\begin{equation}
N_{1}(\tau )=\int n(\mathbf{k},t)d\mathbf{k\simeq }\sum\nolimits_{\mathbf{k}%
}n(\mathbf{k},t)\left( \Delta k\right) ^{3}.
\end{equation}

The normalized second-order correlation function is now defined via 
\begin{align}
g^{(2)}(\mathbf{k},\mathbf{k^{\prime}},t) & =\frac{\langle\hat{a}^{\dagger }(%
\mathbf{k},t)\hat{a}^{\dagger}(\mathbf{k}^{\prime},t)\hat{a}(\mathbf{k}%
^{\prime},t)\hat{a}(\mathbf{k},t)\rangle}{\langle\hat{n}(\mathbf{k}%
,t)\rangle\langle\hat{n}(\mathbf{k}^{\prime},t)\rangle}  \notag \\
& =\frac{\langle:\hat{n}(\mathbf{k},t)\hat{n}(\mathbf{k}^{\prime},t):\rangle 
}{\langle\hat{n}(\mathbf{k},t)\rangle\langle\hat{n}(\mathbf{k}^{\prime
},t)\rangle},  \label{g2-densities}
\end{align}
which is equivalent to Eq.~(\ref{g2}), provided that $\hat{a}(\mathbf{k},t)$
and $\hat{a}_{\mathbf{k}}(t)$ are defined on the same momentum-space lattice.

Finally, we define the normalized variance of the particle number difference 
$\hat{n}_{\mathbf{k}}(t)-\hat{n}_{\mathbf{k}^{\prime }}(t)$,%
\begin{equation}
V_{v}(\mathbf{k},\mathbf{k}^{\prime },t)=1+\frac{\langle :\left[ \Delta
\left( \hat{n}_{\mathbf{k}}(t)-\hat{n}_{\mathbf{k}^{\prime }}(t)\right) %
\right] ^{2}:\rangle }{\langle \hat{n}_{\mathbf{k}}(t)\rangle +\langle \hat{n%
}_{\mathbf{k}^{\prime }}(t)\rangle },  \label{Var-cont}
\end{equation}%
where the particle number operators are to be understood according to Eq.~(%
\ref{n-cont}). The subscript $v$ in Eq.~(\ref{Var-cont}) signifies the fact
that in the continuous model describing an infinite system, the variance is
defined for the counting volume $v=\left( \Delta k\right) ^{3}$, and
therefore Eq. (\ref{Var-cont}) is equivalent to Eq.~(\ref{variance}). 
Accordingly, the results for $V_{v}(\mathbf{k},\mathbf{k}^{\prime },t)$
depend on this volume; unlike the pair correlation function $g^{(2)}(\mathbf{%
k},\mathbf{k^{\prime }},t)$, the scaling of the numerator and of the
denominator with respect to $\left( \Delta k\right) ^{3}$ does not lead to
cancellation. The explicit dependence on $\left( \Delta k\right) ^{3}$
becomes evident if we rewrite Eq.~(\ref{Var-cont}) via the densities $\hat{n}%
(\mathbf{k},t)$, which are independent of $\left( \Delta k\right) ^{3}$: 
\begin{equation}
V_{v}(\mathbf{k},\mathbf{k}^{\prime },t)=1+\left( \Delta k\right) ^{3}\frac{%
\langle :\left[ \Delta \left( \hat{n}(\mathbf{k},t)-\hat{n}(\mathbf{k}%
^{\prime },t)\right) \right] ^{2}:\rangle }{\langle \hat{n}(\mathbf{k}%
,t)\rangle +\langle \hat{n}(\mathbf{k}^{\prime },t)\rangle } .
\label{Var-alt}
\end{equation}%
As we see, for a given negative value of the normally ordered variance $%
\langle :\left[ \Delta \left( \hat{n}(\mathbf{k},t)-\hat{n}(\mathbf{k}%
^{\prime },t)\right) \right] ^{2}:\rangle $, the degree of squeezing below
the shot noise level degrades ($V_{v}$ becomes closer to one) as the
counting volume is decreased.

Specific numerical results obtained for $V_{v}(\mathbf{k},\mathbf{k}^{\prime
},t)$ should not be confused with the requirement that physical observables
should not depend on the choice of the lattice spacing $\Delta k$. Rather,
the calculated variance for a given computational grid corresponds
physically to fluctuations in the difference of particle \emph{numbers} in
the lattice volume element $(\Delta k)^{3}$. For a given density, the number
of particles in a smaller volume is lower and the correlations between
number-difference fluctuations becomes weaker. Stronger correlation signals
can be obtained via the procedure of binning (see below), in which case the
counting volume is chosen to be larger than $(\Delta k)^{3}$.

To further clarify this point we note that the unnormalized variance of the
continuous \emph{density}-difference fluctuations, 
for $\mathbf{k} \neq \mathbf{k}^{\prime }$,%
\begin{align}
& \langle \left[ \Delta \left( \hat{n}(\mathbf{k},t)-\hat{n}(\mathbf{k}%
^{\prime },t)\right) \right] ^{2}\rangle =[\langle \hat{n}(\mathbf{k}%
,t)\rangle +\langle \hat{n}(\mathbf{k}^{\prime },t)\rangle ]\delta ^{(3)}(0)
\notag \\
& +\left\langle :\left[ \Delta \left( \hat{n}(\mathbf{k},t)-\hat{n}(\mathbf{k%
}^{\prime },t)\right) \right] ^{2}:\right\rangle ,  \label{Var-density}
\end{align}%
includes (after rewriting it in terms of the normally ordered operator
products) a term proportional to a delta-function $\delta ^{(3)}(0)$, which
is the shot-noise level. While strong quantum correlation and the reduction
of density-difference fluctuations below the shot-noise level is generically
due to negative values of the normally ordered variance, the \emph{degree}
of squeezing below the delta-function shot-noise is not well defined via
Eq.~(\ref{Var-density}). On the other hand, operating with the variance of
fluctuations between the particle \emph{numbers}, as in Eq.~(\ref{Var-cont}%
), makes the definition of the shot-noise unambiguous and therefore the
degree of squeezing is well defined. Albeit, the squeezing now depends on
the counting volume $\left( \Delta k\right) ^{3}$.

\subsection{Parameter values}

Taking a $^{87}$Rb ($m_{1}=1.43\times 10^{-25}$ kg, $m_{0}=2m_{1}$)
experiment as our specific example, we have performed simulations for the
following set of parameters. The initial molecular BEC density was chosen to
be $\rho _{0}=5\times 10^{19}$ m$^{-3}$ for the uniform system. For the
nonuniform systems, the same value was chosen to be the molecular BEC peak
density. The size of the uniform system $L$ was the same in all three
spatial dimensions, with $L=1.38\times 10^{-5}$ m, and the lattice grid
contained $64^{3}$ points. Thus, the lattice spacing in momentum space was $%
\Delta k=2\pi /L=4.56\times 10^{5}$ m$^{-1}$, while the maximum cutoff
momentum was $k_{\max }=1.46\times 10^{7}$ m$^{-1}$, which is about twice
the resonant momentum $k_{0}=$ $\sqrt{2m_{1}|\Delta _{\mathrm{eff}}|/\hbar }$
and exceeds the physically relevant range of momenta. The total initial
number of molecules was $N_{0}(0)=\rho _{0}L^{3}=1.3\times 10^{5}$ in all
uniform cases. (The values for the nonuniform system are given in Sec. %
\ref{nonuniform}.)

The atom-molecule coupling strength was $\chi =7\times 10^{-7}$ m$^{3/2}/$s.
The bare detuning $\Delta $ in different cases was chosen to result in the
same initial effective detuning $\Delta _{\mathrm{eff}}=\Delta
+(U_{01}-U_{00}/2)\rho _{0}=-1.96\times 10^{4}$ s$^{-1}$ irrespective of the
mean-field interaction contributions. The values of $s$-wave scattering
lengths for the molecule-molecule, atom-molecule, and atom-atom interactions
are given in the respective subsection below. In all simulations, the number
of stochastic trajectories for calculating the expectation values was $7500$
(unless stated otherwise) and the time step was $\Delta t=1.26\times 10^{-6}$
s. Most of the simulations are performed for durations $t/t_{0}=2.5$, where $%
t_{0}=(\chi \sqrt{\rho _{0}})^{-1}=0.2$ ms is the time scale.

With a spatially uniform molecular field and no depletion or two-body
interactions, we validated our code by comparing the numerical
results with those of the analytically soluble model, Eq.~(\ref%
{Heisenberg-eqs}). The numerical code for solving the positive-$P$
stochastic differential equations was implemented using the XMDS software
package \cite{xmds}. Within the sampling errors, the simulated momentum
space mode populations $n_{\mathbf{k}}(t)=n(\mathbf{k},t)\left( \Delta
k\right) ^{3}$ and the correlation functions were in excellent agreement
with the analytic results given by Eqs.~(\ref{n-moment}), (\ref{m-moment})
and (\ref{analytic correlation functions}). The subsections below extend
these results to include the molecular depletion, $s$-wave scattering
interactions, and the treatment of nonuniform systems.

\subsection{Role of the molecular field depletion}

\label{depletion}

\subsubsection{Total particle numbers and atom-atom recombination}

\begin{figure}[tbp]
\includegraphics[width=5.3cm]{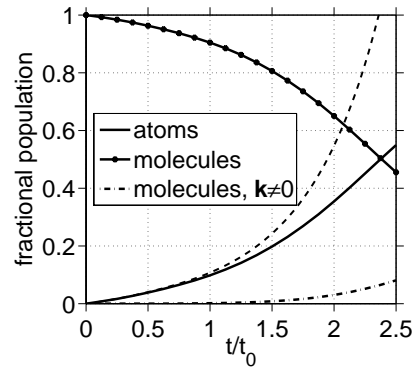}
\caption{Fractional population in the molecular and atomic fields, $%
N_{0}(t)/N_{0}(0)$ and $N_{1}(t)/2N_{0}(0)$, as a function of time $t$. The
time is in units of $t_{0}=1/g=1/(\protect\chi \protect\sqrt{\protect\rho %
_{0}})$, which in this example is $t_{0}=0.2$ ms. The dash-dotted line is
the fraction of molecules in the noncondensate modes, $\mathbf{k}\neq 0$.
The dashed line is the analytic result for the total atom number in the
undepleted molecular field approximation.}
\label{Figure1}
\end{figure}

Here, we simulate Eqs.~(\ref{Positive-P-eqs}) for a uniform system and all $%
s $-wave scattering interactions set to zero ($a_{ij}=0$, and hence $%
\Delta=\Delta_{\mathrm{eff}}$). Including the molecular field in the
dynamics leads to its depletion and a corresponding reduction in the number
of atoms produced compared to the analytically soluble case.

In Fig.~\ref{Figure1} we plot the total fractional number of molecules and
atoms, $N_{0}(t)/N_{0}(0)$ and $N_{1}(t)/2N_{0}(0)$, as a function of time.
The dashed line shows the total atom number in the undepleted molecular
field approximation, Eq.~(\ref{total-atom-number}), for comparison. We see
that the undepleted molecular-field result agrees well with the exact
quantum result for time durations resulting in the conversion of less than $%
10\%$ molecules. In the present example, $10\%$ conversion corresponds to $%
2.6\times 10^{4}$ atoms produced [$N_{1}(t)/2N_{0}(0)\simeq 0.1$] and occurs
at $t/t_{0}\simeq 1$. At this time the discrepancy between the exact
numerical and the approximate analytic results is $\sim 9\%$. For longer
times the discrepancy increases and the approximate analytic result
eventually produces an exponentially growing output. This unphysical
behavior is an obvious consequence of the fact that the undepleted molecular
field approximation is no longer valid.

Physically, the total number of atoms produced saturates due to the
depletion of the molecular condensate containing a finite number of
molecules to start with. At the same time, the dynamics of dissociation --
after the initial spontaneous regime -- is affected by the reverse process of
recombination of atom pairs into molecules. The dashed-dotted line in %
Fig.~\ref{Figure1} shows the total number of molecules outside the condensate
mode $\mathbf{k}=0$. Since the dissociation in this uniform system starts
with all molecules being initially in the condensate mode, the population of
the $\mathbf{k}\neq 0$ modes can only occur (in the absence of elastic
collisions, $a_{ij}=0$) due to atom-atom recombination. Note that the
recombination can in principle involve atom pairs with any momenta, and not
necessarily those with equal and opposite momenta. This has the effect of
reducing atom-atom correlations in the long time limit (see below), hence
the term \textquotedblleft rogue association\textquotedblright\ which we
will use.

The population of the molecular noncondensate modes implies effective
heating of the molecular gas and can be regarded as the reverse of the rogue
dissociation known to occur in the opposite process of association of an
atomic BEC into a molecular BEC \cite{Holland,rogue-dissociation-JJ}. In the
example of Fig.~\ref{Figure1}, $18\%$ of the remaining molecules at $%
t/t_{0}\simeq 2.5$ are outside the condensate mode. At the same time, the
total number of molecules remaining is about $45\%$ of the initial number.

\subsubsection{Atomic momentum distribution}

\begin{figure}[tbp]
\hspace{0.5cm}\includegraphics[width=6.3cm]{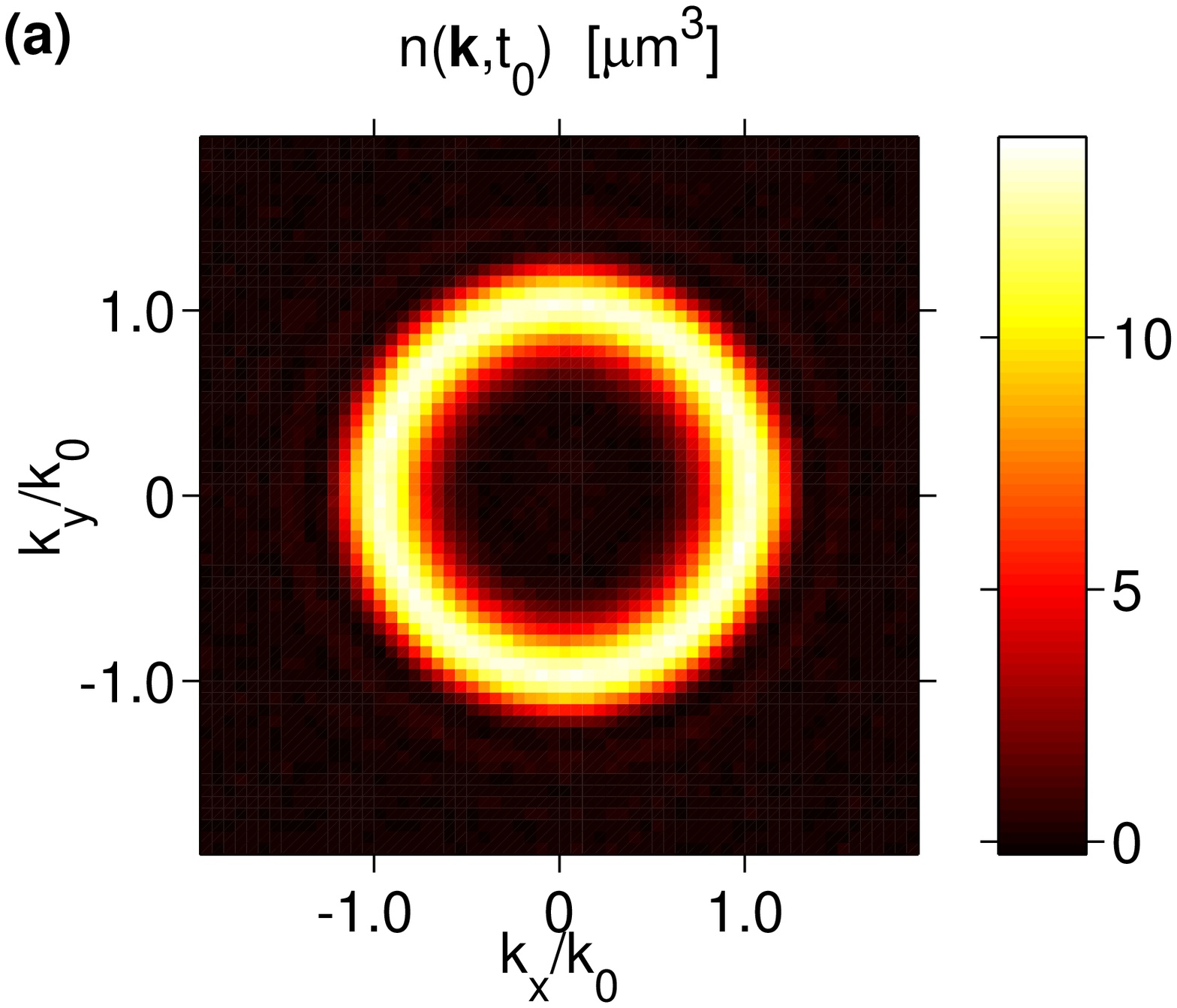} %
\includegraphics[width=5.9cm]{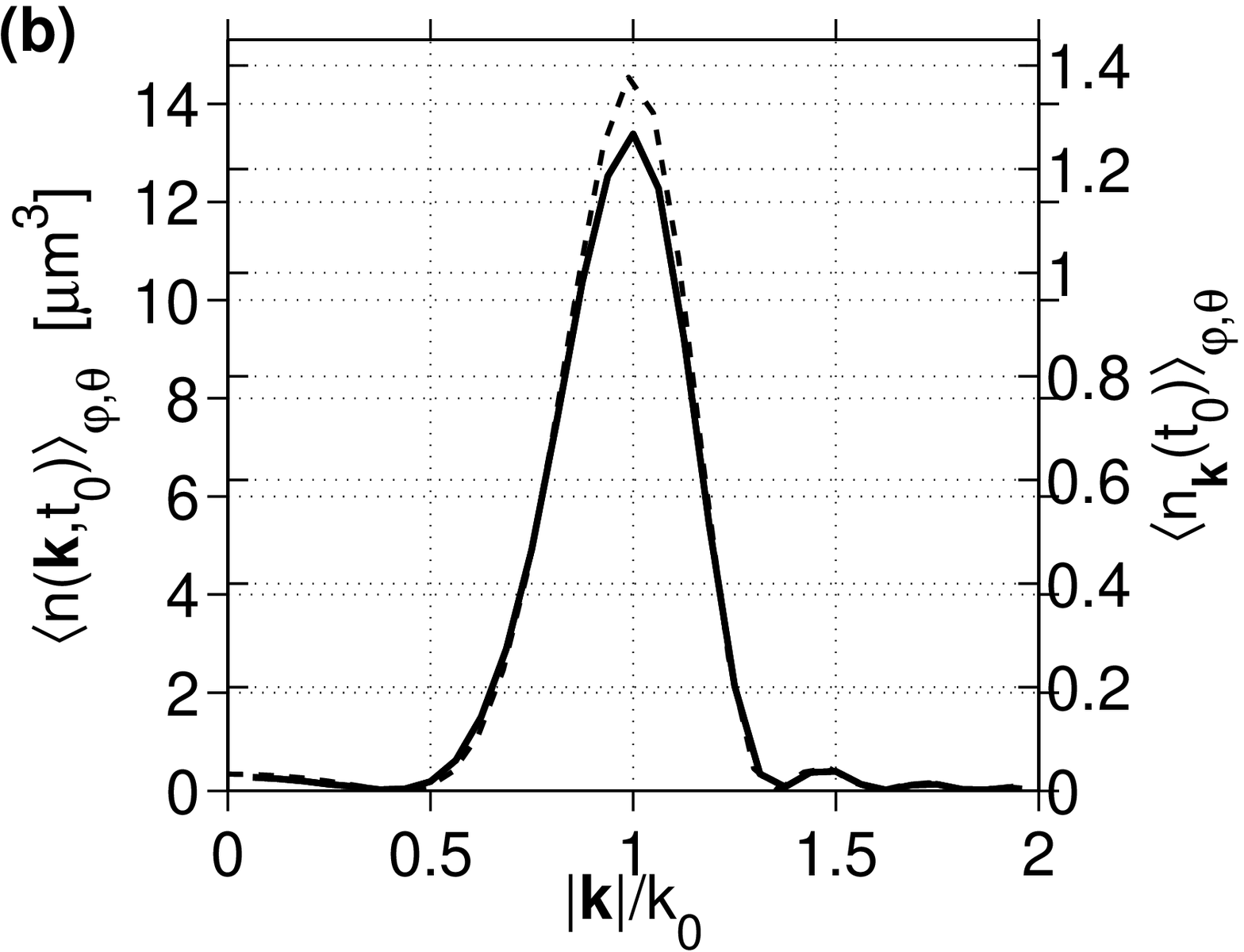}
\caption{(a) (color online) Slice through the origin of the 3D atomic density in momentum
space $n(\mathbf{k},t)$ [in units of $\protect\mu $m$^{3}$] at $t=t_{0}$,
where $t_{0}=1/(\protect\chi \protect\sqrt{\protect\rho _{0}})=0.2$ ms. The
momentum components $k_{x}$ and $k_{y}$ are in units of $k_{0}=$ $\protect%
\sqrt{2m_{1}|\Delta _{\mathrm{eff}}|/\hbar }$, which in this example is $%
k_{0}=7.3\times 10^{6}$ m$^{-1}$. (b) Angle-averaged density distribution $%
\langle n(\mathbf{k},t)\rangle _{\protect\varphi ,\protect\theta }$ and mode
occupations $\langle n_{\mathbf{k}}(t)\rangle _{\protect\varphi ,\protect%
\theta }=$ $\langle n(\mathbf{k},t)\rangle _{\protect\varphi ,\protect\theta %
}(\Delta k)^{3}$ as a function of the absolute momentum $|\mathbf{k}|$ at $%
t=t_{0}$ (solid line). The lattice spacing is $\Delta k=4.55\times 10^{5}$ m$%
^{-1}$. The dashed line is the result for the analytically soluble case with
the undepleted molecular field, Eq. (\protect\ref{n-moment}).}
\label{Figure2}
\end{figure}

In Fig.~\ref{Figure2}(a) we plot a slice through the origin of the atomic
momentum density distribution $n(\mathbf{k},t)$ at $t/t_{0}=1$. We see a
clear ring structure corresponding to the distribution of atoms around the
surface of a sphere of radius $k_{0}$. Similar ring structures have been
observed in the time-of-flight spatial column densities of atoms produced in
dissociation experiments using $^{87}$Rb$_{2}$ and $^{40}$K$_{2}$ molecules 
\cite{Durr,Greiner}.

To simplify the presentation and comparison of the results we make use of
the spherical symmetry of the problem and also plot angle-averaged
distributions $\langle n(\mathbf{k},t)\rangle _{\varphi ,\theta }$ and $%
\langle n_{\mathbf{k}}(t)\rangle _{\varphi ,\theta }$ [see Fig. \ref{Figure2}(b)] 
as a function of the absolute momentum $|\mathbf{k}|$. Here, the brackets 
$\langle \ldots \rangle _{\varphi ,\theta }$ refer to the procedure of
averaging of the quantum mechanical expectation values over the angles $%
\varphi $ and $\theta $ in a spherical coordinate system. 

Comparison of the exact numerical result for the angle-averaged population
distribution $\langle n_{\mathbf{k}}(t)\rangle _{\varphi ,\theta }$ and the
respective analytic result, Eq.~(\ref{n-moment}), shows that the discrepancy
at $t/t_{0}=1$ is less than $8\%$, with the peak value of $\langle n_{%
\mathbf{k}}(t)\rangle _{\varphi ,\theta }$ at $|\mathbf{k}|/k_{0}=1$ being
reduced from the undepleted result of $\sinh ^{2}(1)=1.38$ to $1.27$. \ For
longer time durations the discrepancy increases. The small oscillatory
structure seen in the wings of the distribution functions (for both the
exact numerical and analytic curves) is additional evidence that the
sampling error due to stochastic averaging is small. This oscillatory
behavior is characteristic of the highly detuned modes outside the spherical
shell as these modes do not experience an exponential growth but rather
oscillate at the spontaneous level. In terms of the approximate analytic
solutions, the oscillations take place for the modes with $k=|\mathbf{k}|$
for which the gain coefficient $g_{k}\equiv \lbrack g^{2}-\Delta
_{k}^{2}]^{1/2}$ is pure imaginary. This occurs when $\Delta _{k}^{2}=[\hbar 
\mathbf{k}^{2}/(2m_{1})+\Delta _{\mathrm{eff}}]^{2}>g^{2}$, in which case
the $\sinh ^{2}(g_{k}t)=\sinh ^{2}(i|g_{k}|t)$ term in Eq. (\ref{n-moment})
turns into $\sin ^{2}(|g_{k}|t)$ thus producing the oscillations.

\subsubsection{Atom-atom correlations}

\begin{figure}[tbp]
\hspace{0.1cm}\includegraphics[width=6.3cm]{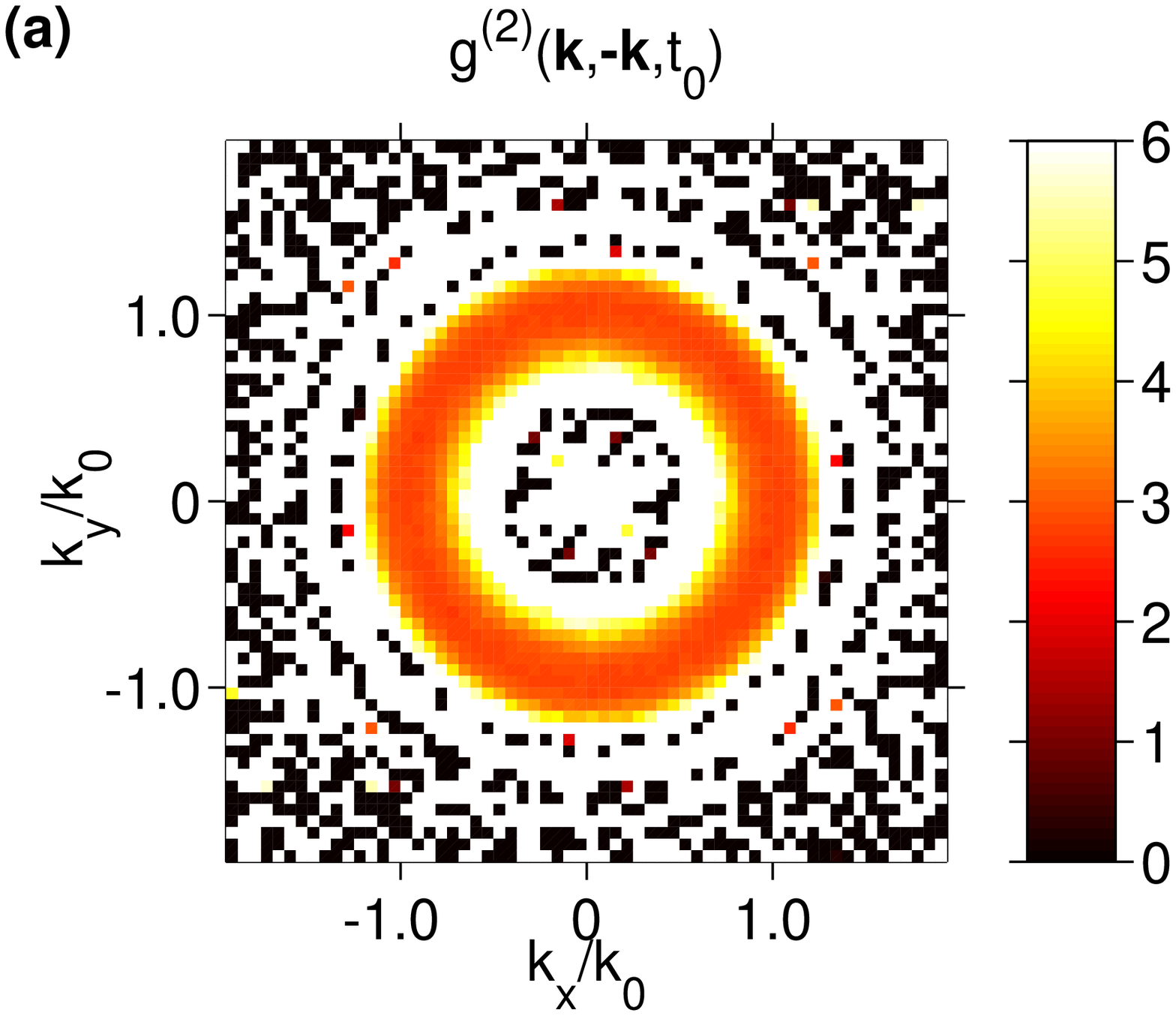} %
\includegraphics[width=6.1cm]{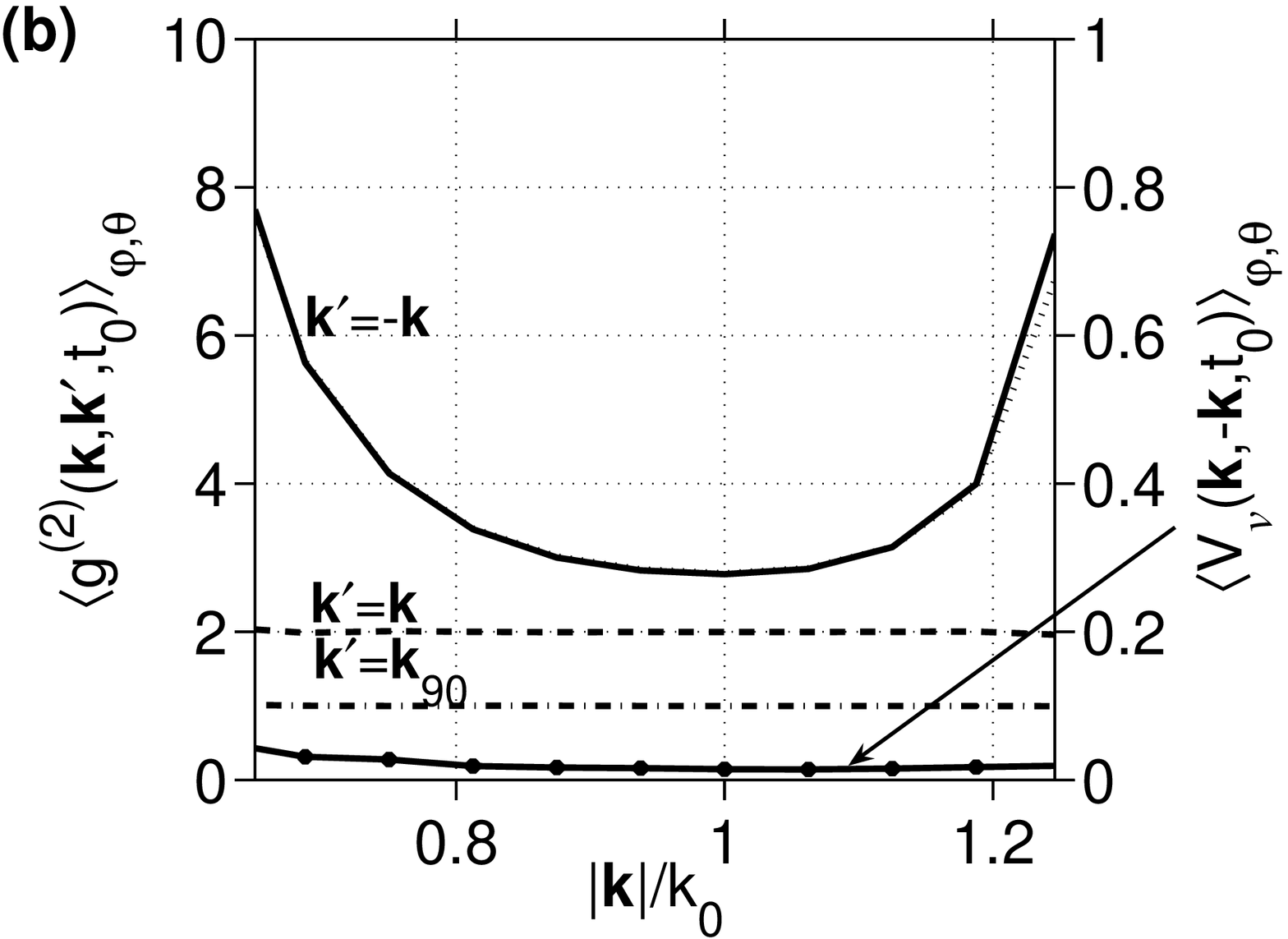}
\caption{(a) (color online) Slice through the origin of the 3D correlation function $%
g^{(2)}(\mathbf{k},-\mathbf{k},t)$ at $t=t_{0}$. The corresponding density
profile is shown in Fig.~\protect\ref{Figure2}(a). The sampling errors due to
stochastic averaging are small only within the region of $0.65<|\mathbf{k}%
|/k_{0}<1.25$, where the correlation function ranges between $\sim 2.8$ and $%
6$. (b) Angle-averaged correlation functions $\langle g^{(2)}(\mathbf{k},%
\mathbf{k}^{\prime },t)\rangle _{\protect\varphi ,\protect\theta }$ (for $%
\mathbf{k}^{\prime }=-\mathbf{k}$, $\mathbf{k}$, $\mathbf{k}_{90}$) at $%
t=t_{0}$ as a function of the absolute momentum $|\mathbf{k}|$ within the
spherical shell $0.65<|\mathbf{k}|/k_{0}<1.25$. The solid line with the dots
(right scale) is the respective angle-averaged variance $\langle V_{v}(%
\mathbf{k},-\mathbf{k},t)\rangle _{\protect\varphi ,\protect\theta }$.}
\label{Figure3}
\end{figure}

The pair correlation function for the atoms with opposite momenta, $g^{(2)}(%
\mathbf{k},-\mathbf{k},t)$, is shown in Fig. \ref{Figure3}(a). Here, we plot a
slice through the origin of the 3D correlation function at $t/t_{0}=1$. Due
to the finite number of stochastic trajectories and the normalization of the
correlation function to the product of mode occupancies, the stochastic
ensemble average gives highly noisy results in the regions of $\mathbf{k}$%
space where the mode populations are small ($n_{\mathbf{k}}\lesssim 0.3$).
This is seen within the central ($|\mathbf{k}|/k_{0}<0.65$) and outer ($|%
\mathbf{k}|/k_{0}>1.25$) region of the slice plot where the sampling error
is large \cite{Comment1}. Within the spherical shell of $0.65<|\mathbf{k}%
|/k_{0}<1.25$, on the other hand, the mode populations are higher and the
sampling errors are negligible. Within this region the results for $g^{(2)}(%
\mathbf{k},-\mathbf{k},t)$ scale according to the colormap shown on the
right of Fig. \ref{Figure3}(a).

Figure \ref{Figure3}(b) shows the angle-averaged correlation functions $%
\langle g^{(2)}(\mathbf{k},\mathbf{k}^{\prime },t)\rangle _{\varphi ,\theta
} $ within the spherical shell $0.65<|\mathbf{k}|/k_{0}<1.25$, at $t/t_{0}=1$%
. As expected, for $\mathbf{k}^{\prime }=-\mathbf{k}$ the correlations are
super bunched, $\langle g^{(2)}(\mathbf{k},-\mathbf{k},t)\rangle _{\varphi
,\theta }>2$, while for $\mathbf{k}^{\prime }=\mathbf{k}$ we see the thermal
level of bunching $\langle g^{(2)}(\mathbf{k},\mathbf{k},t)\rangle _{\varphi
,\theta }=2$ characteristic of Gaussian statistics. The line showing an
uncorrelated level of $\langle g^{(2)}(\mathbf{k},\mathbf{k}_{90},t)\rangle
_{\varphi ,\theta }=1$ is for the pair of momenta $\mathbf{k}$ and $\mathbf{k%
}^{\prime }=\mathbf{k}_{90}$, where $\mathbf{k}_{90}$ corresponds to a $%
90^{\circ }$ degree rotation of the density distribution in the ($%
k_{z,}k_{y} $) plane around the $k_{z}$ axis.

Apart for the fact that the pump depletion results in slightly decreased
mode populations relative to the analytic result, the correlations follow
closely Eqs.~(\ref{analytic correlation functions}), except that $n_{\mathbf{%
k}}(t)$ is now the actual, decreased, mode population. This observation is valid for
at least the simulated time window of $t/t_{0}=2.5$, corresponding to about $%
55$\% conversion. This is expected since the correlations are due to
momentum conservation, which is unaffected by the decreasing rate of
conversion, which is the major effect of the molecular depletion. Thus, we
find that the pair correlation for equal and opposite momenta is still well
approximated by 
\begin{equation}
\langle g^{(2)}(\mathbf{k},-\mathbf{k},t)\rangle _{\varphi ,\theta }\simeq
2+1/\langle n_{\mathbf{k}}(t)\rangle _{\varphi ,\theta }.  \label{g-approx}
\end{equation}%
Since $\langle n_{\mathbf{k}}(t)\rangle _{\varphi ,\theta }$ is slightly
less than in the undepleted molecular field
approximation, the pair correlation Eq.~(\ref{g-approx}) is accordingly
higher. In Fig.~\ref{Figure3}(b) the result of Eq.~(\ref{g-approx}) is plotted
by the dotted line and is almost invisible as it follows closely the actual
numerical value calculated using Eq.~(\ref{g2-densities}).

The fact that the correlations can be approximated by Eq.~(\ref{g-approx})
shows that Wick factorization continues to approximately hold for time
durations corresponding to $\sim 55\%$ conversion. We recall that the
ingredients of this result are the vanishing off-diagonal normal moments, $%
\langle \hat{a}_{\mathbf{k}}^{\dagger }(t)\hat{a}_{-\mathbf{k}}(t)\rangle =0$%
, and the relationship of Eq.~(\ref{m-n relation}) between the normal and
anomalous moments. These ingredients continue to hold in the present example.

\begin{figure}[tbp]
\includegraphics[width=5.7cm]{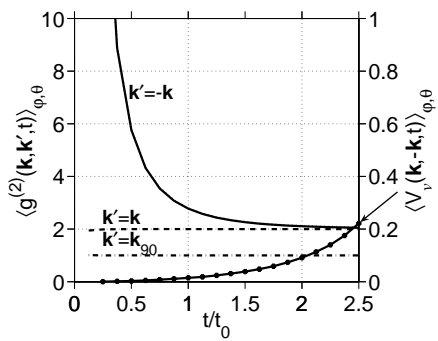}
\caption{Dynamics of the angle-averaged pair correlation functions $\langle
g^{(2)}(\mathbf{k},\mathbf{k}^{\prime },t)\rangle _{\protect\varphi ,\protect%
\theta }$ at $|\mathbf{k}|/k_{0}=1$, for $\mathbf{k}^{\prime }=-\mathbf{k}$, 
$\mathbf{k}$, and $\mathbf{k}_{90}$. The time is in the same units as in
Figs. \protect\ref{Figure1} and \protect\ref{Figure2}. The solid line with
the dots (right scale) is the variance of the number-difference fluctuations for equal but
opposite momenta, $\langle V_{v}(\mathbf{k},-\mathbf{k},t)\rangle _{\protect%
\varphi ,\protect\theta }$, at $|\mathbf{k}|/k_{0}=1$.}
\label{Figure4}
\end{figure}

Figure \ref{Figure4} shows the temporal behavior of the angle-averaged pair
correlation functions $\langle g^{(2)}(\mathbf{k},\mathbf{k}^{\prime
},t)\rangle _{\varphi ,\theta }$ at $|\mathbf{k}|/k_{0}=1$ corresponding to
the peak momentum in the density distribution. The approximate result
for $\langle g^{(2)}(\mathbf{k},-\mathbf{k},t)\rangle _{\varphi ,\theta }$,
given by Eq.~(\ref{g-approx}), is plotted with a dotted line, but is coincident 
with the actual numerical value shown by the solid line.

\subsubsection{Number-difference variance}

\label{subsubsection-variance}

As we mentioned earlier, the pair correlation function becomes a less
sensitive measure of the correlation strength when the mode occupation
numbers increase. In particular, as $n_{\mathbf{k}}(t)$ increases with time
the pair correlation approaches $g^{(2)}(%
\mathbf{k},-\mathbf{k},t)\rightarrow 2$ which coincides with the level of
thermal bunching for the autocorrelation function. In this regime, the
variation in $g^{(2)}(\mathbf{k},-\mathbf{k},t)$ is no longer significant.
As a result, distinguishing between strong and weak correlations becomes
problematic.

A more sensitive measure of the correlation strength is provided by the
variance of the particle number-difference fluctuations between the
correlated modes, Eq.~(\ref{variance}). 
The angle-averaged number-difference variance for the modes with
equal but opposite momenta, $\langle V_{v}(\mathbf{k},-\mathbf{k},t)\rangle
_{\varphi ,\theta }$, is shown in Figs. \ref{Figure3}(b) and \ref{Figure4} by
solid lines with dots. As we see from Fig.~\ref{Figure3}(b), the variance is
suppressed well below the shot-noise level $\langle V_{v}(\mathbf{k},-%
\mathbf{k},t)\rangle _{\varphi ,\theta }<1$ for the entire spectral range.

Similarly, the variance $\langle V_{v}(\mathbf{k},-%
\mathbf{k},t)\rangle _{\varphi ,\theta }$ shows
sub-shot-noise fluctuations in the entire simulated time window (see Fig.~\ref{Figure4}).
It increases above the ideal result of $\langle V_{v}(\mathbf{k},-%
\mathbf{k},t)\rangle _{\varphi ,\theta }=0$ towards the end of the simulated
time window. In particular, at $t/t_{0}=2.5$ it is given by $\langle V_{v}(%
\mathbf{k},-\mathbf{k},t)\rangle _{\varphi ,\theta }\simeq 0.22 $ which
corresponds to $78\%$ squeezing below the shot-noise level. The reduction of
squeezing is attributed to atom-atom recombination into molecules with
nonzero momenta, $\mathbf{k}\neq 0$ (see Fig.~\ref{Figure1}). Since this
process involves pairs of atoms with nonopposite momenta, 
a certain fraction of the atoms is left without their
correlated partners of opposite momentum. As a result, the overall strength
of the correlation signal for the atoms with equal and opposite momenta is
reduced.

The reduction of the correlation strength due to atom-atom recombination
becomes a sizeable effect after dissociation times ($t/t_{0}\gtrsim 2$)
corresponding to more than $\sim 35$\% conversion. At very short times (less
than $10$\% conversion) our exact numerical results confirm that the
predictions of the analytically soluble model with the undepleted molecular
field are a good approximation. At longer times the correct treatment of the
system requires the inclusion of the molecular field depletion.

As shown in Ref.~\cite{Jack-Pu}, the molecular depletion can be taken into
account using a pairing mean-field method which is simpler to implement
numerically than the present (exact) positive-$P$ method. In the mean-field
method, the molecules are described as a coherent field in the condensate
mode $\langle \hat{b}_{\mathbf{k}=0}(t)\rangle $ while the atoms are treated
at the level of diagonal normal and anomalous populations, $\langle \hat{a}_{%
\mathbf{k}}^{\dagger }(t)\hat{a}_{\mathbf{k}}(t)\rangle $ and $\langle \hat{a%
}_{\mathbf{k}}(t)\hat{a}_{-\mathbf{k}}(t)\rangle $. While imposing Wick's
factorization scheme and assuming perfect correlation between the atoms with
opposite momenta at all times, this treatment does not allow for atom-atom
recombination into molecules outside the condensate mode. Accordingly, the
predictions of the theory are a good approximation only for dissociation
durations resulting in no more than $\sim 35$\% conversion, as can be seen
from the comparison with the present exact results.

The pairing mean-field method performs better in the long time
limit if the only observables of interest are the total particle numbers.
Its performance can be expected to improve for
calculating the correlation functions if it is expanded to include the
molecular noncondensate modes and the off-diagonal terms in the atomic
normal and anomalous densities.

Before turning to the analysis of atom correlations in the presence of $s$%
-wave scattering interactions, we note that the positive-$P$ simulations
without the two-body collisions fail for durations longer than $%
t/t_{0}\simeq 2.5$. At this time about $55\%$ of molecules have been
converted into atoms. The failure of the simulations for longer durations is
due to the generic boundary term problem that occurs in the positive-$P$
method \cite{Gilchrist,DeuarDrummond}. Despite this, our results are a
significant improvement and extend beyond the regime of validity of the
simple undepleted molecular field approximation or the pairing mean-field
theory. A possible route to increase the simulation time is to use
stochastic gauges \cite{DeuarDrummond,Stochastic-gauges}, which is, however,
beyond the scope of the present paper.

\subsection{Role of the $s$-wave scattering interactions}

\label{two-body-terms}

\begin{table}[tp]
\begin{tabular}{|c|c|c|}
\hline
$a_{ij}$ (nm) &{ \rule{0ex}{2.5ex} $U_{ij}$ ($\times 10^{-17}$ m$^{3}$/s) }& $t_{\mathrm{sim}%
}/t_{0}$ ($t_{0}=0.2$ ms) \\ \hline
$a_{00}=2$ & $U_{00}=0.921$ & 2.125 \\ \hline
$a_{01}=2$ & $U_{01}=1.38$ & 1.75 \\ \hline
$a_{11}=2$ & $U_{11}=1.84$ & 1.25 \\ \hline
$a_{ij}=2$ & $U_{ij}$ -- as above & 1.25 \\ \hline
$a_{11}=5.3$ & $U_{11}=4.88$ & 0.5 \\ \hline
\end{tabular}%
\caption{Scattering lengths $a_{ij}$, coupling constants $U_{ij}$, and the
respective simulated time duration $t_{\mathrm{sim}}$ (in units of $t_{0}=1/(%
\protect\chi \protect\sqrt{\protect\rho _{0}})=0.2$ ms) for simulations with
the $s$-wave scattering interaction terms. For each case, the values of the
remaining terms were set to zero, i.e., $a_{ij}=0$ except for the quoted
term. All other parameters are as in Sec. IV A.}
\label{two-body}
\end{table}

Here, in addition to simulating the molecular field depletion we include the 
$s$-wave scattering interaction terms. Table \ref{two-body} shows the
different cases considered, together with the simulated time window $t_{%
\mathrm{sim}}$. The simulations are shorter than in the absence of the $s$%
-wave scattering since the positive-$P$ boundary term problem \cite%
{Gilchrist,DeuarDrummond} becomes more severe for quartic interactions.
The simulation times were chosen to be less than those for which the sampling error was observed to grow rapidly  \cite{DeuarDrummond}.

The shortest $t_{\mathrm{sim}}$ is for the case of $a_{11}=5.3$ nm which is
the natural background scattering length for $^{87}$Rb atoms. Longer $t_{%
\mathrm{sim}}$ are achieved with smaller values of $a_{ij}$. The scattering
length can in principle be tuned during dissociation by applying a magnetic
field in the vicinity of a Feshbach resonance. In this case the dissociation
itself would have to be invoked independently, using optical Raman
transitions. While this is unlikely to simultaneously result in the specific
values of $a_{ij}$ quoted in Table \ref{two-body}, our main goal is the
understanding of the role of each physical process separately -- by
quantitatively estimating their relative importance and the potential
disruptive effect on atom-atom correlations.

The results of simulations of the cases of Table \ref{two-body} show that
the momentum distribution and the pair correlation for atoms with opposite
momenta remain (within the simulated time windows) very close to the values
obtained with no $s$-wave interactions. For example, in the case of $%
a_{11}=2 $ nm, the peak occupancy reduces from $1.70$ to $1.62$ at $%
t/t_{0}=1.25$, while the corresponding pair correlation still follows Eq.~(%
\ref{g-approx}). The reduction in the mode population is attributed to the
mean-field energy shift due to atom-atom interactions, which dynamically
changes the phase matching condition and hence results in a drift of the
effective detuning $\Delta_{\mathrm{eff}}$ as the populations evolve. As a
result, the resonant momentum also drifts causing less efficient conversion
into the initially resonant mode. Nevertheless, since the bare detuning $%
|\Delta |$ in our examples is still much larger than all mean-field phase
shifts $U_{ij}\langle \hat{\Psi}_{j}^{\dagger }(\mathbf{x},t)\hat{\Psi}_{j}(%
\mathbf{x},t)\rangle $, the resonant momentum is still well approximated by $%
k_{0}\simeq $ $\sqrt{2m_{1}|\Delta |/\hbar }$.

The expected reduction of the actual correlation strength due to $s$-wave
collisions and the resulting redistribution of the atomic momenta is best
revealed through the variance of the number-difference fluctuations, $%
\langle V_{v}(\mathbf{k},-\mathbf{k},t)\rangle _{\varphi ,\theta }$. For
example, in the case of $a_{ij}=2$ nm (which is the worst case, among the
simulations surviving up to durations of $t/t_{0}=1.25$), the variance at
the resonant momentum $k_{0}$ increases from the collisionless result of $%
0.024$ to $0.032$ at $t/t_{0}=1.25$. This implies the reduction of
number-difference squeezing from $97.6$\% to $96.8$\%. The overall effect
is, however, smaller than decorrelation due to atom-atom recombination
discussed earlier, which in this example is responsible for the reduction of
squeezing from $100$\% down to $97.6$\%.

\subsection{Nonuniform systems and mode mixing}

\label{nonuniform}

We now turn to the analysis of atom-atom correlations in nonuniform systems
corresponding to dissociation of realistic trapped BECs. The net effect of
inhomogeneity is mode mixing, which can dramatically affect the
correlations. The quantitative details are best understood in the undepleted
molecular field approximation. In this case, the atom-atom recombination and 
$s$-wave scattering interactions are absent, and any reduction in atom-atom
correlation strength -- compared to the uniform case -- is due to
mode mixing.

Hence, the molecular field amplitude $\Psi _{0}(\mathbf{x},0)=\sqrt{\rho
_{0}(\mathbf{x})}$ [where $\rho _{0}(\mathbf{x})$ is the density] can be
absorbed into an effective coupling $g(\mathbf{x})\equiv \chi \sqrt{\rho
_{0}(\mathbf{x})}$. This leads to the same equations of motion for the
atomic field operator $\hat{\Psi}_{1}(\mathbf{x},t)$ as Eq. (\ref%
{Heisenberg-eqs}),\ except that the coupling constant $g$ is now a function
of $\mathbf{x}$. Introducing the Fourier transform $\tilde{g}(\mathbf{k}%
)=(2\pi )^{-3/2}\int d\mathbf{x}g(\mathbf{x})\exp (-i\mathbf{k}\cdot \mathbf{%
x})$, the equation of motion for the atomic Fourier component $\hat{a}(%
\mathbf{k},t)$ is%
\begin{eqnarray}
\frac{d\hat{a}(\mathbf{k},t)}{dt} &=&-i\left( \frac{\hbar \mathbf{k}^{2}}{%
2m_{1}}+\Delta _{\mathrm{eff}}\right)  \hat{a}(\mathbf{k},t) 
\notag \\
&&+\frac{1}{(2\pi )^{3/2}}\int d\mathbf{k}^{\prime }\tilde{g}(\mathbf{k}+%
\mathbf{k}^{\prime })\hat{a}^{\dag }(\mathbf{k}^{\prime },t).
\label{Heisenberg-nonuniform}
\end{eqnarray}%
As we see, $\hat{a}(\mathbf{k},t)$ couples to the range of momentum
components $\mathbf{k}^{\prime }$ in the conjugate field $\hat{a}^{\dag }(%
\mathbf{k}^{\prime },t)$, which we refer to as mode mixing. This is in
contrast to the uniform case, where $\tilde{g}(\mathbf{k}+\mathbf{k}^{\prime
})$ is the delta function $\tilde{g}(\mathbf{k}+\mathbf{k}^{\prime })=(2\pi
)^{3/2}g\delta (\mathbf{k}+\mathbf{k}^{\prime })$, so that $\hat{a}(\mathbf{k%
},t)$ couples only to the conjugate field at the opposite momentum $\hat{a}%
^{\dag }(-\mathbf{k},t)$.

The overall effect of the mode mixing is to correlate the atoms with
momentum $\mathbf{k}$ not only with the opposite momentum $-\mathbf{k}$, but
also with the atoms distributed in a range of momenta around $-\mathbf{k}$.
As a result, for a given strength of $\tilde{g}(\mathbf{k})$ the pair
correlation between $\mathbf{k}$ and $-\mathbf{k}$ is expected to be reduced
compared to the uniform case.

To characterize mode mixing quantitatively we simulate the positive-$P$
stochastic differential equations that are equivalent to Eqs.~(\ref%
{Heisenberg-nonuniform}). The initial molecular BEC is assumed to be formed
in a spherically symmetric harmonic trap with equal trap oscillation
frequencies $\omega \equiv \omega _{x,y,z}$. It is
assumed to be in a coherent state, with the density distribution given by the
Thomas-Fermi parabola $\rho _{0}(\mathbf{x})=$ $\rho _{0}(0)(1-|\mathbf{x|}%
^{2}/R_{TF}^{2})$ for $|\mathbf{x|}<R_{TF}$, and $\rho _{0}(\mathbf{x})=0$
for $|\mathbf{x|\geq }R_{TF}$. Here, $\rho _{0}(0)$ is the peak density,
while $R_{TF}=\sqrt{2\hbar U_{00}\rho _{0}(0)/(m_{0}\omega ^{2})}$ is the
Thomas-Fermi (TF) radius.

\subsubsection{Case 1: weak inhomogeneity}

The choice of parameter values in this set of simulations is specifically
targeted to give a situation which is directly comparable with the results
of the previous uniform system. The uniform results for a cubic box of side $%
L_{u}$ can be regarded as an approximation to a realistic nonuniform system
if $L_{u}$ is matched with the characteristic size $\sim 2R_{TF}$ of the
initial molecular BEC, while the uniform density $\rho _{0}$ is matched with
the peak density $\rho _{0}(0)$. More specifically, we choose $R_{TF}=(8\pi
/15)^{-1/3}L_{u}$ \ and $\rho _{0}(0)=\rho _{0}$, which gives the same
initial total number of molecules in both cases, $N_{0}(0)=(8\pi /15)\rho
_{0}(0)R_{TF}^{3}=\rho _{0}L_{u}^{3}$.

\begin{figure*}[ptb]
~~~~~~~~~~~\includegraphics[height=7.1cm]{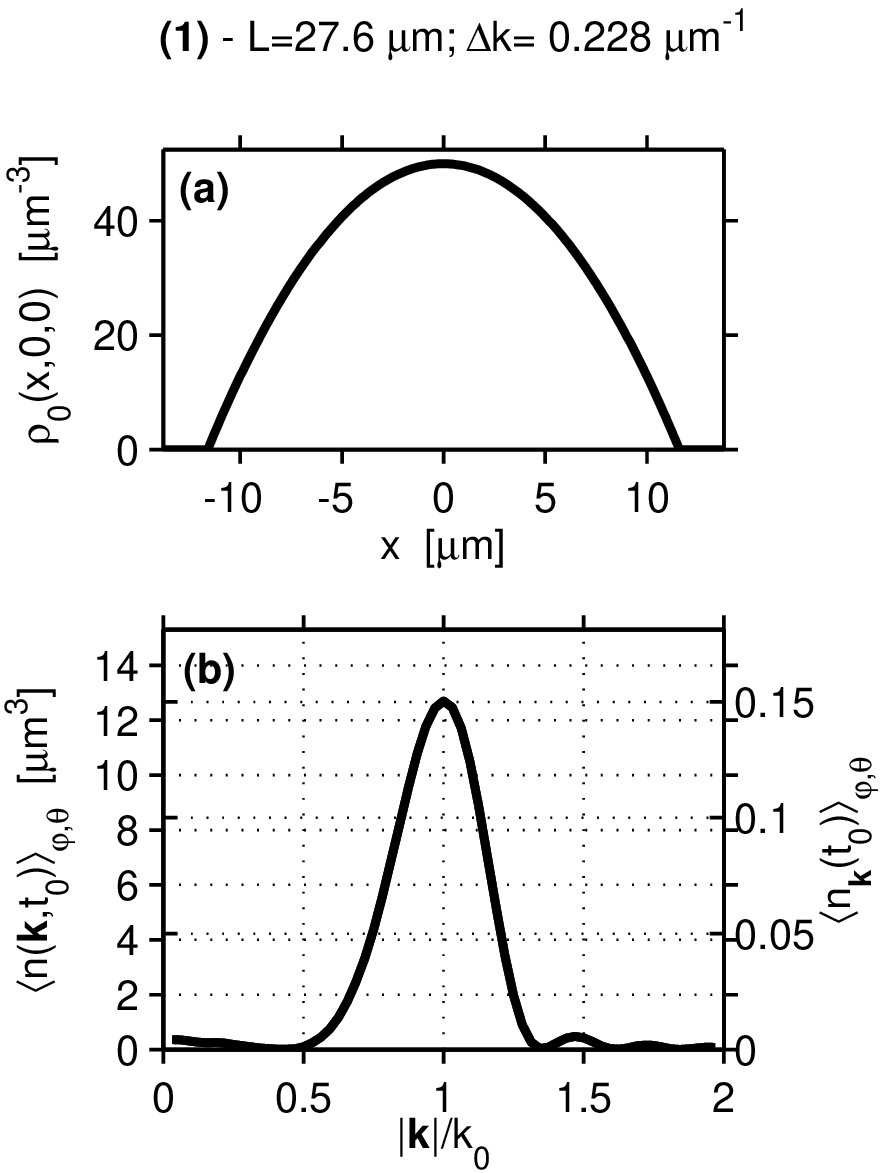}~~~~~\includegraphics[height=7.1cm]{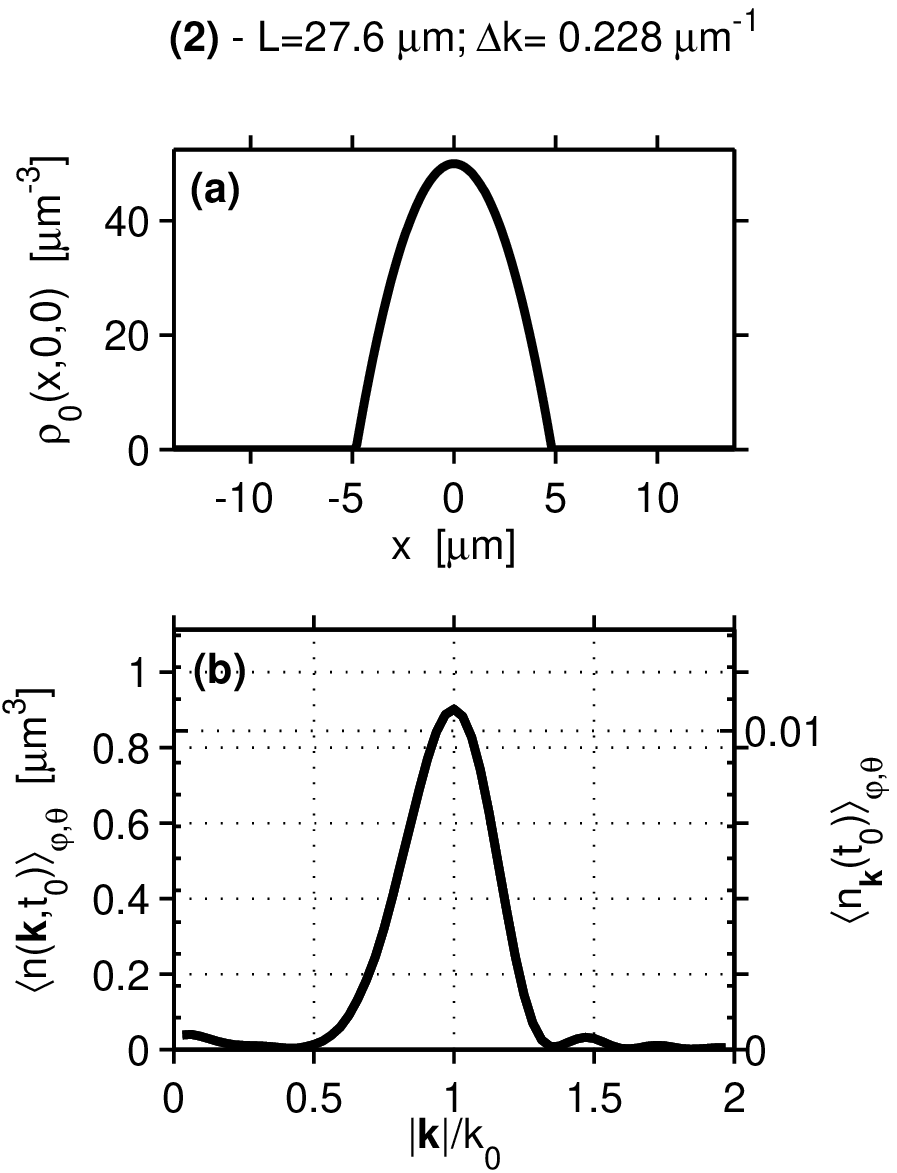}~~~~\includegraphics[height=7.1cm]{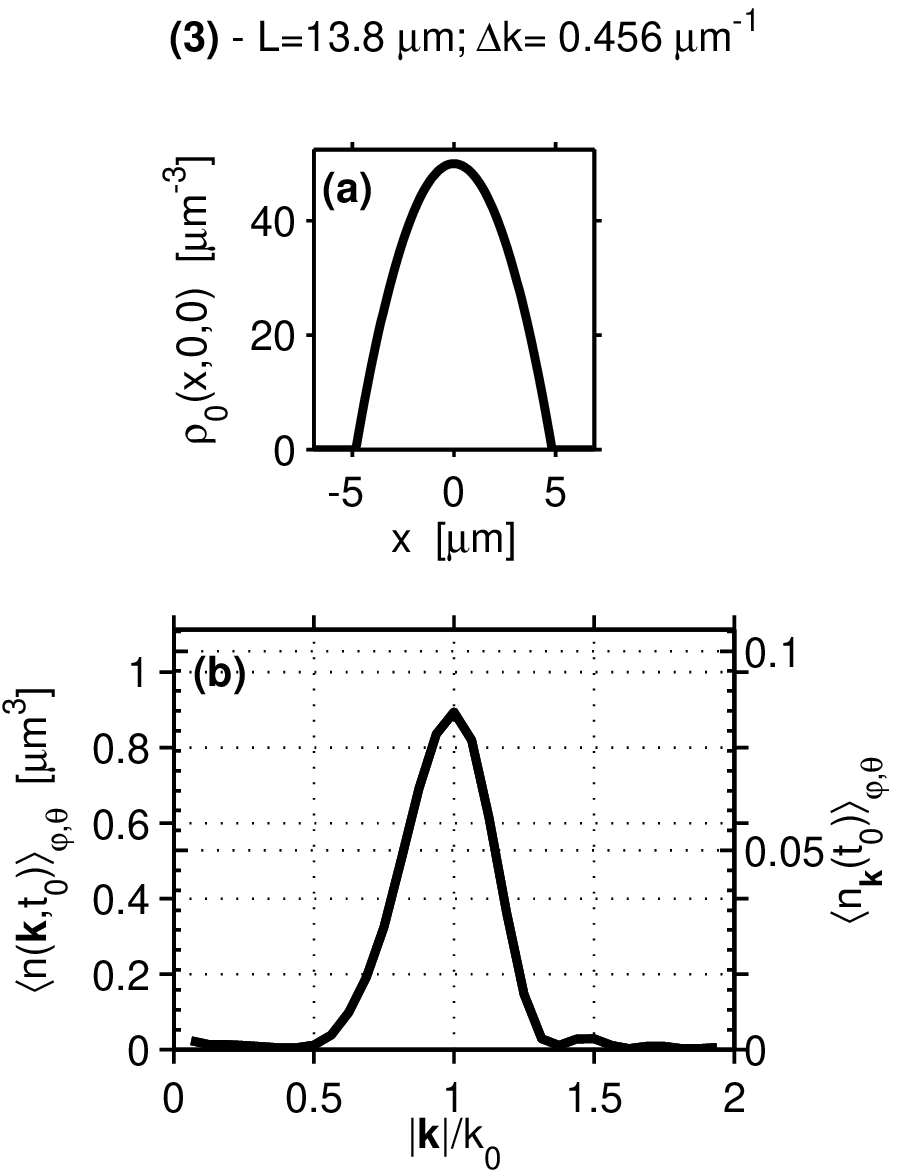}
\includegraphics[width=4.8cm]{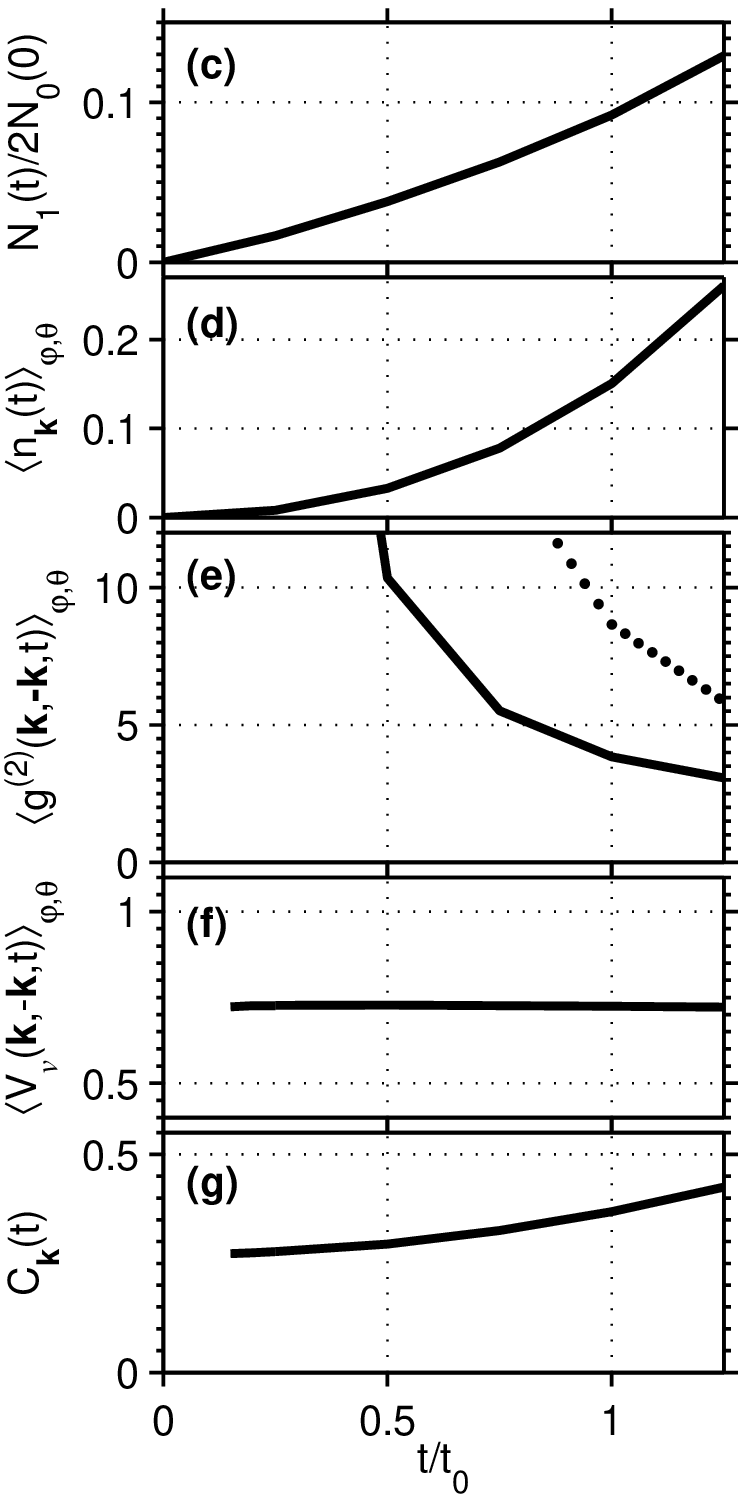}~~~~~~~~\includegraphics[width=4.95cm]{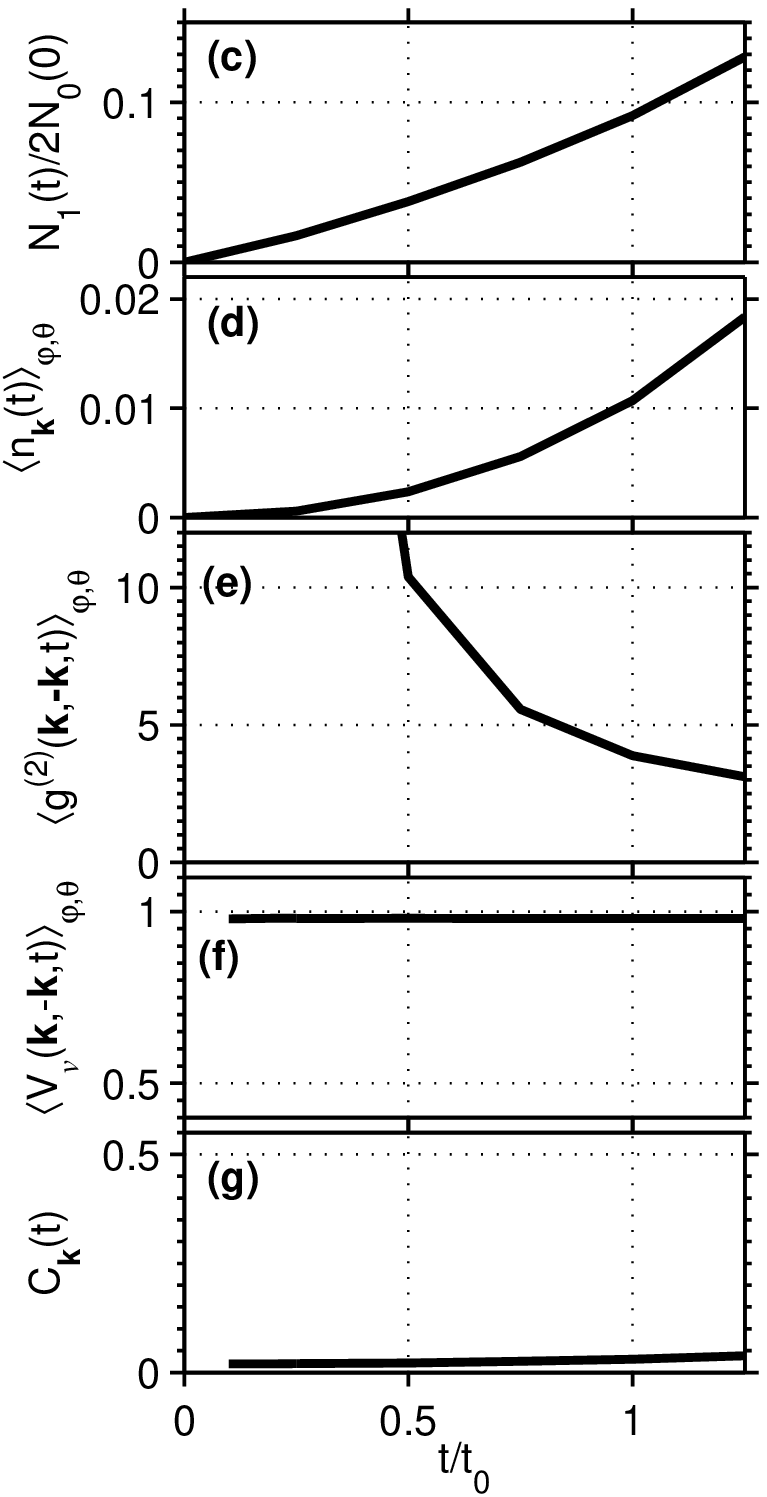}~~~~~~~~~\includegraphics[width=4.8cm]{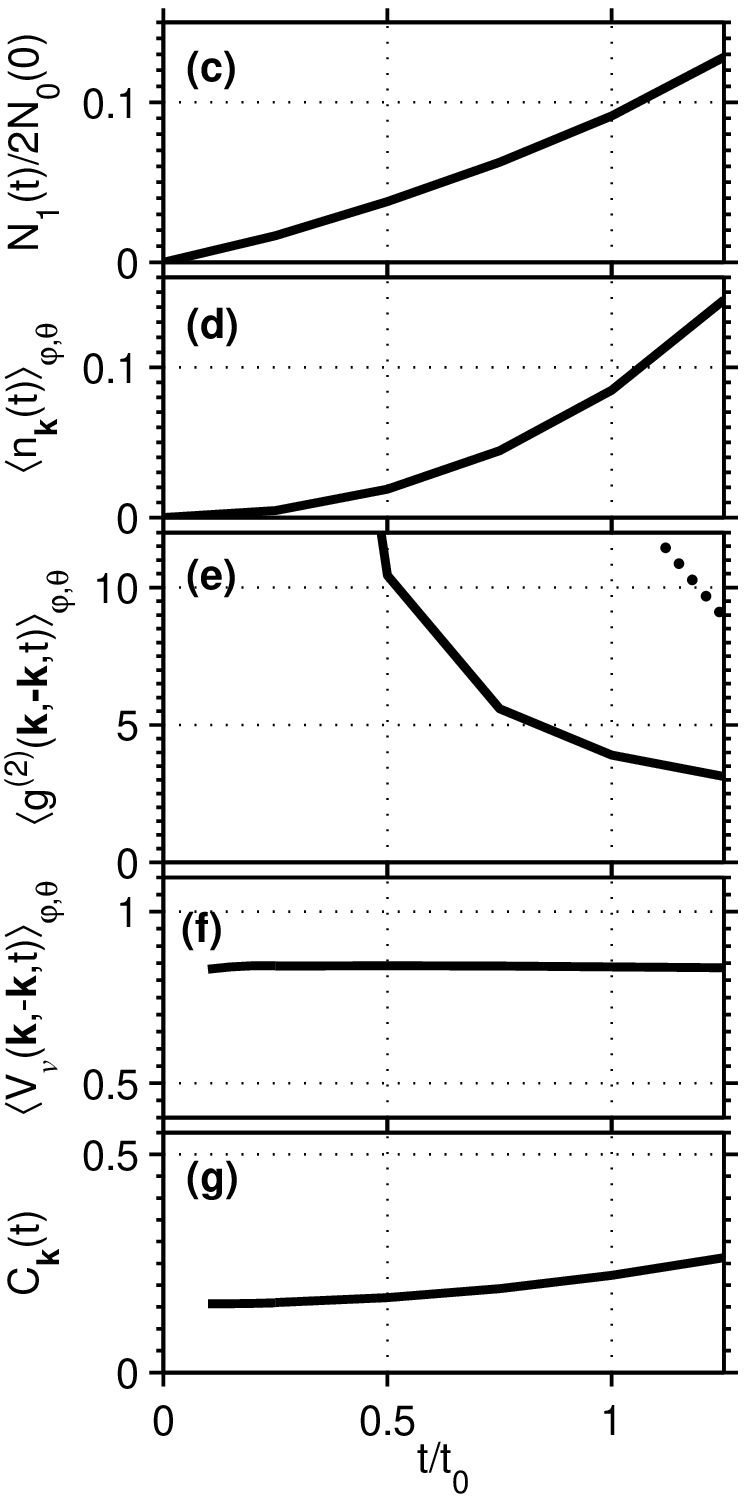}
\caption{Comparison between the results of simulations of weakly inhomogeneous
($R_{TF}=11.6$ $\mu$m) and strongly inhomogeneous ($R_{TF}=4.82$ $\mu$m)
systems, columns 1 and 2, having the same peak density $\rho_{0}(0)%
=5\times10^{19}$ m$^{-3}$. Column 3 refers to the same physical system as in
column 2, except that the length of the simulated spatial domain $L$ is halved. 
Note that the lines are provided as guides-to-the-eye between calculated data points. 
(a) Slice through the origin of the initial molecular BEC density
profile, $\rho_{0}(x,0,0)$. (b) Angle-averaged momentum density $\langle
n(\mathbf{k},t)\rangle_{\varphi,\theta}$ and distribution of mode occupations
$\langle n_{\mathbf{k}}(t)\rangle_{\varphi,\theta}$ at $t=t_{0}$ as
a function of the absolute momentum $|\mathbf{k}|/k_{0}$, where $k_{0}%
=7.3\times10^{6}$ m$^{-1}$ as in Fig. \ref{Figure2}. (c) Fraction of the total
number of dissociated atoms relative to $2N_{0}(0)$ vs time. (d)
Angle-averaged mode population $\langle n(\mathbf{k},t)\rangle_{\varphi
,\theta}$ vs time. Note the different scales. (e) Second-order 
pair correlation function $\langle g^{(2)}%
(\mathbf{k},-\mathbf{k},t)\rangle_{\varphi,\theta}$ vs time. The dotted 
line shows the idealized result
$\langle g^{(2)}_{\max}(\mathbf{k},-\mathbf{k},t)\rangle_{\varphi,\theta}=2+1/%
\langle n_{\mathbf{k}}(t)\rangle_{\varphi,\theta}$ for comparison.
(f) Number-difference variance $\langle V_{v}(\mathbf{k},-\mathbf{k}%
,t)\rangle_{\varphi,\theta}$ vs time.  (g) Correlation coefficients $C_{\mathbf{k}}(t)$ at $|\mathbf{k}%
|=k_{0}$ vs time. %
The quantities shown in (d)-(g) are at $|\mathbf{k}|=k_{0}$ and
the time is in units of $t_{0}=1/(\chi\sqrt{\rho_{0}(0)%
})=0.2$ ms in all cases. The number of stochastic trajectories 
used for ensemble averaging is 2000.}%
\label{FigureX}%
\end{figure*}%

Thus, in the present example (with the results shown in Fig. \ref{FigureX},
column 1) the peak density is $\rho _{0}(0)=5\times 10^{19}$ m$^{-3}$, while
the TF radius is $R_{TF}=1.16\times 10^{-5}$ m, which would correspond to a
molecule-molecule scattering length of $a_{00}=2$ nm in a relatively weak
trap with oscillation frequency $\omega /2\pi =8$ Hz. This gives the total
initial number of molecules $N_{0}(0)=1.3\times 10^{5}$. The simulations are
performed on the spatial domain of $L=2.76\times 10^{-5}$ m in each
dimension, which is twice the size of the earlier uniform system $L_{u}$.
The lattice grid contained $128^{3}$ points. Thus, the lattice spacing in
momentum space is now $\Delta k=2\pi /L=2.28\times 10^{5}$ m$^{-1} $ (half
that in the uniform system), while the maximum cutoff momentum is the same $%
k_{\max }=1.46\times 10^{7}$ m$^{-1}$.

From Fig.~\ref{FigureX} (column 1) we see that the atomic density in
momentum space (b) and the total number of atoms (c) follow closely the
predictions of the size-matched uniform system shown in Figs. \ref{Figure2}%
(b) and \ref{Figure1}(a). At the same time, the mode
populations are about eight times smaller, which is simply due to the
fact that the simulation of the nonuniform system is performed within a
spatial domain which is twice as large, $L=2L_{u}$. 
Accordingly, the respective counting volumes $%
(\Delta k)^{3}=(2\pi /L)^{3}$ and $(\Delta k_{u})^{3}=(2\pi /L_{u})^{3}$ 
are different
by a factor of eight, so that the approximately equal densities $n(\mathbf{k}%
,t)\simeq n_{u}(\mathbf{k},t)$ give different mode populations $n_{\mathbf{k}%
}(t)=n(\mathbf{k},t)(\Delta k)^{3}$ and $n_{\mathbf{k}}^{(u)}(t)=n_{u}(%
\mathbf{k},t)(\Delta k_{u})^{3}$. Scaling the uniform result for $n_{u}(%
\mathbf{k},t)$ with respect to the counting volume used in
the nonuniform system gives $n_{u}(\mathbf{k},t)(\Delta k)^{3}$ that
approximates well the actual calculated value of $n_{\mathbf{k}}(t)$ shown
in Fig. \ref{FigureX}. 

While the results for the momentum-space density and the atom number are in
good quantitative agreement with the respective results of the size-matched
uniform system, the same is not true for the correlation function $g^{(2)}(%
\mathbf{k},-\mathbf{k},t)$. The angle-averaged pair correlation $\langle
g^{(2)}(\mathbf{k},-\mathbf{k},t)\rangle _{\varphi ,\theta }$ at $|\mathbf{k}%
|=k_{0}$ in the present nonuniform system is shown in Fig.~\ref{FigureX}(e)
by the full line. Before comparing this result with that of the uniform
system [given by the idealized result of $g_{\max }^{(2)}(\mathbf{k},-%
\mathbf{k},t)=2+1/n_{\mathbf{k}}(t)$], we note that the correct
interpretation of the normalized pair correlation is in the excess of the
probability of joint detection of pairs of atoms at $\mathbf{k}$ and $-%
\mathbf{k}$ relative to that probability in an uncorrelated state.
Operationally, this corresponds to determining the number of particles in a
\textquotedblleft detection\textquotedblright\ volume $(\Delta k)^{3}$.
Therefore, the pair correlation $\langle g^{(2)}(\mathbf{k},-\mathbf{k}%
,t)\rangle _{\varphi ,\theta }$ for the nonuniform system can be compared
with the idealized result $g_{\max }^{(2)}(\mathbf{k},-\mathbf{k},t)=2+1/n_{%
\mathbf{k}}(t)$ provided that $n_{\mathbf{k}}(t)$ is evaluated for the same
counting volume $(\Delta k)^{3}$ as in the nonuniform simulation.

The idealized result $g_{\max }^{(2)}(\mathbf{k},-\mathbf{k},t)$ obtained in
this way is shown in Fig.~\ref{FigureX}(e) by the dotted line. We
immediately see that the strength of the pair correlation for the nonuniform
system $\langle g^{(2)}(\mathbf{k},-\mathbf{k},t)\rangle _{\varphi ,\theta }$
is substantially lower than $g_{\max }^{(2)}(\mathbf{k},-\mathbf{k},t)$. For
comparison, the same quantity calculated numerically in the uniform system
was well approximated by the idealized result [see discussions of Figs.~\ref{Figure3}(b) and %
\ref{Figure4}], where it was almost invisible behind the solid line for $%
\langle g^{(2)}(\mathbf{k},-\mathbf{k},t)\rangle _{\varphi ,\theta }$.

We note that the correlation between the atoms with opposite momenta is
still stronger than in an uncorrelated state, $\langle g^{(2)}(\mathbf{k},-%
\mathbf{k},t)\rangle _{\varphi ,\theta }>1$. We have also calculated the
pair correlation functions $\langle g^{(2)}(\mathbf{k},\mathbf{k},t)\rangle
_{\varphi ,\theta }$ and $\langle g^{(2)}(\mathbf{k},\mathbf{k}%
_{90},t)\rangle _{\varphi ,\theta }$. As in the uniform case, these were
given by $2$ and $1$, respectively, implying that the mode mixing has
negligible effect on these correlations.

Another measure of the reduction of the correlation strength between the
atoms with opposite momenta can be obtained using the variance of the
number-difference fluctuations $V_{v}(\mathbf{k},-\mathbf{k},t)$, Eq.~(\ref%
{Var-cont}). The angle-averaged variance $\langle V_{v}(\mathbf{k},-\mathbf{k%
},t)\rangle _{\varphi ,\theta }$ at $|\mathbf{k}|=k_{0}$ as a function of
time $t$ is shown in Fig.~\ref{FigureX}(f), and is given approximately by $%
\langle V_{v}(\mathbf{k},-\mathbf{k},t)\rangle _{\varphi ,\theta }\simeq
0.72 $, where $v=(\Delta k)^{3}$ is the elementary volume element of the
computational grid. This corresponds to $28\%$ squeezing of fluctuations
below the shot-noise level, which is a significant degradation 
compared to the
ideal case of $100\%$ squeezing obtained in the uniform systems in the
absence of molecular depletion or $s$-wave scattering interactions. The
reduction of the correlation strength due to mode mixing is a much stronger
effect than decorrelation due to the atom-atom recombination and $s$-wave
scattering, analyzed in Secs. \ref{depletion} and \ref{two-body-terms}.

Finally, in Fig.~\ref{FigureX}(g) we plot the correlation coefficient%
\begin{equation}
C_{\mathbf{k}}(t)\equiv \frac{|\langle m_{\mathbf{k}}(t)\rangle _{\varphi
,\theta }|^{2}}{\langle n_{\mathbf{k}}(t)\rangle _{\varphi ,\theta
}(1+\langle n_{\mathbf{k}}(t)\rangle _{\varphi ,\theta })}  \label{Ck}
\end{equation}%
as a function of time $t$, at $\mathbf{|k|}=k_{0}$. According to the
inequality (\ref{m-nonideal}), this gives a useful measure of the departure [%
$C_{\mathbf{k}}(t)<1$] of the anomalous mode population from the maximally
correlated, ideal result of Eq.~(\ref{max-m}) corresponding to $C_{\mathbf{k}%
}(t)=1$. In addition, the calculated anomalous population $m_{\mathbf{k}}(t)$
gives us a direct verification of the validity of Wick's factorization,
according to which the pair correlation $g^{(2)}(\mathbf{k},-\mathbf{k},t)$
in the present example can be evaluated as in Eq.~(\ref{Wick-m}). We have
checked that the off-diagonal normal moment $\langle \hat{a}_{\mathbf{k}%
}^{\dagger }(t)\hat{a}_{-\mathbf{k}}(t)\rangle $ gives a negligible
contribution, $|\langle \hat{a}_{\mathbf{k}}^{\dagger }(t)\hat{a}_{-\mathbf{k%
}}(t)\rangle |^{2}/n_{\mathbf{k}}^{2}(t)\ll 1$.

\ \

\subsubsection{Cases 2 and 3: strong inhomogeneity}

Columns 2 and 3 in Fig.~\ref{FigureX} show the results of simulations of a
nonuniform system in a tighter trap, with $\omega /2\pi =19.3$ Hz and $%
R_{TF}=4.82\times 10^{-6}$ m. The total initial number of molecules is $%
N_{0}(0)=9.4\times 10^{3}$, while all other parameters are as in the
previous example. Thus, in the present case the inhomogeneity in
position space is stronger. Conversely, the effective coupling $\tilde{g}(%
\mathbf{k})$ is broader in momentum space and therefore the mode mixing has
a stronger effect on the reduction of correlations between the atom pairs
with opposite momenta.

The results presented in column 2 of Fig.~\ref{FigureX} are obtained on the
same computational grid as in column 1, i.e., using $L=2.76\times 10^{-5}$ m
and $128^{3}$ lattice points. Therefore the quantities depending on the
counting volume $v=(\Delta k)^{3}$ -- through the mode occupation number $n_{%
\mathbf{k}}(t)=n(\mathbf{k},t)(\Delta k)^{3}$ -- are directly comparable.

Apart from the obvious reduction in the momentum space density and total
atom number produced as a function of time, the fraction of atoms relative
to the initial number of molecules is almost the same as in the example of
column 1. The density-density correlation function does not reveal
significant quantitative change either. However, as we explained in the
previous subsection, the quantitative aspect of the reduction of the
correlation strength should be assessed relative to the idealized result for 
$g_{\max }^{(2)}(\mathbf{k},-\mathbf{k},t)=2+1/n_{\mathbf{k}}(t)$. In the
present example, $g_{\max }^{(2)}(\mathbf{k},-\mathbf{k},t)$ is off the
scale of Fig.~\ref{FigureX}(e), signaling a much more dramatic reduction in
the correlation strength than before. This is further seen through the
number-difference variance, $\langle V_{v}(\mathbf{k},-\mathbf{k},t)\rangle
_{\varphi ,\theta }\simeq 0.98$, implying only $2\%$ of squeezing below the
shot noise-level. Similarly, the correlation coefficient $C_{\mathbf{k}}(t)$
is much lower than the ideal result of $C_{\mathbf{k}}(t)=1$.

In column 3, Fig.~\ref{FigureX}, we show the results of simulation of the
same physical system as in column 2 except that the simulation is performed
on a smaller ($64^{3}$) computational grid, with half the length $%
L=1.38\times 10^{-5}$ m. Thus, the elementary volume element $v=(\Delta
k)^{3}$, where $\Delta k=2\pi /L$, is eight times larger and therefore the
results depending on the counting volume $v$ are scaled by a factor of eight.
For example, the number-difference variance is now given by $\langle V_{v}(%
\mathbf{k},-\mathbf{k},t)\rangle _{\varphi ,\theta }\simeq 0.84$ ($16\%$
squeezing), demonstrating that the larger counting volume results in a
stronger correlation signal between the particle number-difference
fluctuations.

\subsubsection{Binning}
\label{sec: binning}

The comparison between cases 2 and 3 as a function of the counting volume $%
(\Delta k)^{3}$ is similar to the procedure of binning except that it is
done implicitly -- via the variation of the length $L$ of the simulated
spatial domain in each dimension. Since the minimum acceptable length must
be larger than the characteristic size of the actual physical system, the
relationship $\Delta k=2\pi /L$ puts an upper bound on the volume of the
elementary bin that can be treated in this implicit way.

\begin{figure}[tbp]
\includegraphics[width=6cm]{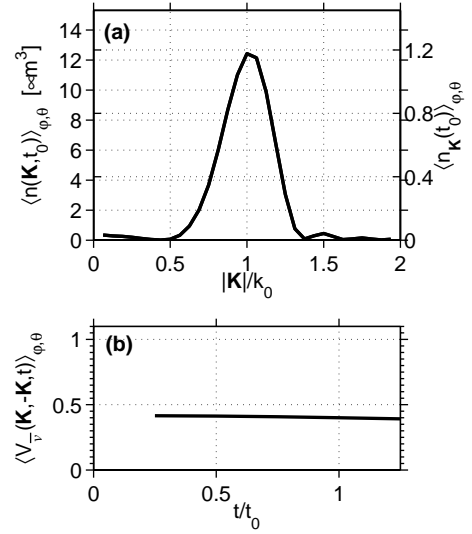}
\caption{(a) Binned distribution of the angle-averaged occupation numbers $%
\langle n_{\mathbf{K}}(t)\rangle _{\protect\varphi ,\protect\theta }$ at $%
t=t_{0}$; and (b) the respective number-difference variance $\langle V_{%
\overline{v}}(\mathbf{K},-\mathbf{K},t)\rangle _{\protect\varphi ,\protect%
\theta }$ at $|\mathbf{K|}=k_{0}$ as a function of time, for the same
physical system as in Fig. \protect\ref{FigureX}, column 1. The graphs can
be compared to those of Figs. \protect\ref{FigureX}(b) and (f), column 1. The $\{%
\mathbf{K\}}$-sublattice of the central momenta of the bins is setup to
correspond to every second lattice point of the original computational grid
in each dimension, so that the bin volume is $\overline{v}=(2\Delta k)^{3}$.
The number of stochastic trajectories simulated is $4500$.}
\label{Figure6}
\end{figure}

To correlate the signals in larger counting volumes, the binning must be
done explicitly.  We introduce the binned number operator
\begin{equation}
\hat{n}_{\mathbf{K}}(t)=\int_{\overline{v}(\mathbf{K})}\hat{n}(\mathbf{k}%
^{\prime },t)d\mathbf{k}^{\prime }\mathbf{\simeq }\sum_{\mathbf{k}\in 
\overline{v}(\mathbf{K})}\hat{n}(\mathbf{k},t)\overline{v}(\mathbf{K}),
\label{binned-nK}
\end{equation}%
which is defined on a $\{\mathbf{K\}}$-sublattice corresponding to the
central momenta of the bins, where $\overline{v}(\mathbf{K})$ is the bin
volume around $\mathbf{K}$.
The variance of fluctuations between the number of particles in different
bins is defined via
\begin{equation}
V_{\overline{v}}(\mathbf{K},\mathbf{K}^{\prime },t)=1+\frac{\langle :\left[
\Delta \left( \hat{n}_{\mathbf{K}}(t)-\hat{n}_{\mathbf{K}^{\prime
}}(t)\right) \right] ^{2}:\rangle }{\langle \hat{n}_{\mathbf{K}}(t)\rangle
+\langle \hat{n}_{\mathbf{K}^{\prime }}(t)\rangle },  \label{binned-variance}
\end{equation}%
where the integration over the bin volume must precede the ensemble
averaging. 

In Fig.~\ref{Figure6} we
present the results of simulations of the same physical system as in %
Fig.~\ref{FigureX}, column 1, except that the results are binned. The bin
counting volume for calculating the occupation numbers $n_{\mathbf{K}%
}(t)=\langle \hat{n}_{\mathbf{K}}(t)\rangle $ and the number-difference
variance $V_{\overline{v}}(\mathbf{K},-\mathbf{K},t)$, Eq.~(\ref%
{binned-variance}) is  $\overline{v}=(2\Delta k)^{3}$. This is
eight times larger than the elementary volume element of the computational
grid $(\Delta k)^{3}$.

As expected, the binned variance shows a stronger correlation signal. The
degree of squeezing increases to $\sim 60\%$ ($\langle V_{\overline{v}}(%
\mathbf{K},-\mathbf{K},t)\rangle _{\varphi ,\theta }\simeq 0.4$, for $|%
\mathbf{K|}=k_{0}$), compared to $28\%$ squeezing in the respective
unbinnned variance of Fig.~\ref{FigureX}(f), column 1. Similar results are obtained
for the cases represented in columns 2 and 3 of Fig.~\ref{FigureX}.

In practice, similar issues arise in time-of-flight images due to the finite
resolution of the imaging system, and binning was employed in the recent
spatial correlation measurements of Ref.~\cite{Greiner}. The correlation
signal between fluctuations of the binned particle numbers was indeed found
to be stronger for larger bins.

\section{Summary}

To summarize, we have performed first-principles quantum dynamical
simulations of dissociation of a BEC of molecular dimers into correlated
atom pairs in three dimensions. The simulations are done using the positive-$%
P$ representation method which allows us to obtain exact results for the
atomic and molecular populations and atom-atom correlations in momentum
space. We have analyzed the effects of molecular depletion, $s$-wave
scattering interactions, and mode mixing in nonuniform condensates.

We find that the most useful measure of the strength of atom-atom
correlations is given by the variance of atom number-difference fluctuations
in a certain range of momenta around the central momenta of interest $%
\mathbf{k}$ and $\mathbf{k}^{\prime }$. The strongest correlation signal
resulting in squeezing of particle number-difference fluctuations below the
shot noise level is obtained for pairs of equal but opposite momenta, $%
\mathbf{k}^{\prime }=-\mathbf{k}$. The degree of squeezing depends on the
geometry of the system and is stronger for BEC samples that are spatially larger 
and more uniform.

The major source of atom-atom decorrelation compared to the ideal result
(achievable in uniform systems and in the absence of molecular depletion) is
mode mixing due to inhomogeneity of the molecular BEC. This suggests that a
preferred experimental strategy to obtain a strong correlation signal is to
start with a large, low density sample. The actual correlation strength
depends also on the counting volume around the central momenta and can be
increased by binning the signals into larger bins.

The next important source of decorrelation is the atom-atom recombination,
which produces increasingly large number of molecules in the initially
unpopulated noncondensate modes as the dissociation proceeds. In our
example the fraction of these molecules is about $20\%$ of the remaining
total number once the overall conversion reaches $50\%$. At this stage, the
reduction in number-difference squeezing due to recombination is about $20\%$%
.

Finally, the simulated $s$-wave scattering interactions are found to have a
less severe effect on atom-atom correlations at least for time durations
corresponding to less than $10\%$ conversion. For longer time durations, the
simulations fail due to the limitations of positive-$P$ method, unless the
atom-atom scattering length is reduced below the typical background value.

\section*{Acknowledgments}

The authors acknowledge stimulating discussions with P. D. Drummond, M. K.
Olsen, J. F. Corney, M. J. Davis, N. Syassen, and T. Volz, as well us the
help provided by the developers of the XMDS software \cite{xmds}, J. Hope,
C. Roy, and T. Vaughan. The research was supported by the Australian
Research Council.

\appendix

\section{Discrete and continuous Fourier transforms}

To treat the uniform system in a cubic box of side $L$ with periodic
boundary conditions, we expand the field operators $\hat{\Psi}_{1}(\mathbf{x}%
,t)$ in terms of the plane-wave momentum modes $\hat{a}_{\mathbf{k}}(t)$: 
\begin{align}
\hat{\Psi}_{1}(\mathbf{x},t)& =\frac{1}{L^{3/2}}\sum\nolimits_{\mathbf{k}}%
\hat{a}_{\mathbf{k}}(t)e^{i\mathbf{k\cdot x}}, \\
\hat{\Psi}_{1}^{\dagger }(\mathbf{x},t)& =\frac{1}{L^{3/2}}\sum\nolimits_{%
\mathbf{k}}\hat{a}_{\mathbf{k}}^{\dagger }(t)e^{-i\mathbf{k\cdot x}}.
\end{align}%
The inverse transforms are%
\begin{align}
\hat{a}_{\mathbf{k}}(t)& =\frac{1}{L^{3/2}}\int_{\mathbf{V}}d\mathbf{x}\;%
\hat{\Psi}_{1}(\mathbf{x},t)e^{-i\mathbf{k\cdot x}}, \\
\hat{a}_{\mathbf{k}}^{\dagger }(t)& =\frac{1}{L^{3/2}}\int_{\mathbf{V}}d%
\mathbf{x}\;\hat{\Psi}_{1}^{\dagger }(\mathbf{x},t)e^{i\mathbf{k\cdot x}}.
\end{align}%
The mode creation and annihilation operators satisfy the usual commutation
relation $[\hat{a}_{\mathbf{k}}(t),\hat{a}_{\mathbf{k}^{\prime }}^{\dagger
}(t)]=\delta _{\mathbf{k},\mathbf{k}^{\prime }}$, where $\delta _{\mathbf{k},%
\mathbf{k}^{\prime }}$ is the Kroenecker delta function, $\mathbf{k}%
=(k_{x},k_{y},k_{z})$ is the momentum (in wave-number units), and $%
k_{i}=(2\pi /L)n_{i}$ ($i=x,y,z,$ $n_{i}=0,\pm 1,\pm 2,\ldots $).

The treatment of the infinite system in free space corresponds to taking the
limit of $L\rightarrow \infty $ ($\Delta k=2\pi /L\rightarrow 0$), together
with converting the sums $\sum_{\mathbf{k}}\equiv \sum_{\mathbf{n}%
}=\sum_{n_{x}}\sum_{n_{y}}\sum_{n_{z}}$ into integrals according to $%
\sum_{n_{i}}\rightarrow \int dn_{i}=(L/2\pi )\int dk_{i}$. The corresponding
continuous Fourier transforms are defined according to 
\begin{align}
\hat{\Psi}_{1}(\mathbf{x},t)& =\frac{1}{(2\pi )^{3/2}}\int d\mathbf{k}\;\hat{%
a}(\mathbf{k},t)e^{i\mathbf{k\cdot x}}, \\
\hat{\Psi}_{1}^{\dagger }(\mathbf{x},t)& =\frac{1}{(2\pi )^{3/2}}\int d%
\mathbf{k}\;\hat{a}^{\dagger }(\mathbf{k},t)e^{-i\mathbf{k\cdot x}},
\end{align}%
with the inverse transforms of 
\begin{align}
\hat{a}(\mathbf{k},t)& =\frac{1}{(2\pi )^{3/2}}\int d\mathbf{x}\;\hat{\Psi}%
_{1}(\mathbf{x},t)e^{-i\mathbf{k\cdot x}}, \\
\hat{a}^{\dagger }(\mathbf{k},t)& \equiv \left[ \hat{a}(\mathbf{k},t)\right]
^{\dagger }=\frac{1}{(2\pi )^{3/2}}\int d\mathbf{x}\;\hat{\Psi}_{1}^{\dagger
}(\mathbf{x},t)e^{i\mathbf{k\cdot x}}, \\
\hat{a}(-\mathbf{k},t)& =\frac{1}{(2\pi )^{3/2}}\int d\mathbf{x}\hat{\Psi}%
_{1}(\mathbf{x},t)e^{i\mathbf{k\cdot x}}, \\
\hat{a}^{\dagger }(-\mathbf{k},t)& \equiv \left[ \hat{a}(-\mathbf{k},t)%
\right] ^{\dagger }=\frac{1}{(2\pi )^{3/2}}\int d\mathbf{x}\hat{\Psi}%
_{1}^{\dagger }(\mathbf{x},t)e^{-i\mathbf{k\cdot x}}.
\end{align}%
The Fourier components $\hat{a}(\mathbf{k})$ and $\hat{a}^{\dagger }(\mathbf{%
k},t)$ satisfy the commutation relation $[\hat{a}(\mathbf{k}),\hat{a}%
^{\dagger }(\mathbf{k}^{\prime })]=\delta (\mathbf{k}-\mathbf{k}^{\prime })$.

When represented on a discrete computational lattice, the continuous Fourier
components are approximated by 
\begin{equation}
\hat{a}(\mathbf{k},t)=\hat{a}_{\mathbf{k}}(t)/\left( \Delta k\right) ^{3/2},
\label{cont-dicrete}
\end{equation}%
and the continuous delta function by 
\begin{equation}
\delta (\mathbf{k}-\mathbf{k}^{\prime })=\delta _{\mathbf{k},\mathbf{k}%
^{\prime }}/\left( \Delta k\right) ^{3},
\end{equation}%
where $\Delta k=2\pi /L$ is the mode spacing in each dimension. The Fourier
components $\hat{a}(\mathbf{k})$ (unlike the mode annihilation operators $%
\hat{a}_{\mathbf{k}}$) have units; $n(\mathbf{k},t)=\langle \hat{a}^{\dagger
}(\mathbf{k},t)\hat{a}(\mathbf{k},t)\rangle $ corresponds to the particle
number density in momentum space, while $n_{\mathbf{k}}(t)=\langle \hat{a}_{%
\mathbf{k}}^{\dagger }(t)\hat{a}_{\mathbf{k}}(t)\rangle $ is the number of
particles in the mode $\mathbf{k}$.

\section{Fourier transforms of the stochastic fields}

Here, we outline the correspondences between the Fourier transforms of the
positive-$P$ stochastic fields and those for the field operators. This is
important for correct implementation of the routines aimed at the 
calculation of various operator moments in Fourier space.

We recall that the two independent stochastic ($c$ number) fields $\Psi _{1}(%
\mathbf{x},t)$ and $\Phi _{1}(\mathbf{x},t)$ are chosen here to correspond
to the annihilation and creation operators $\hat{\Psi}_{1}(\mathbf{x},t)$
and $\hat{\Psi}_{1}^{\dagger }(\mathbf{x},t)$ according to the following
operator correspondences: 
\begin{align}
\hat{\Psi}_{1}(\mathbf{x},t)& \rightarrow \Psi _{1}(\mathbf{x},t), \\
\hat{\Psi}_{1}^{\dagger }(\mathbf{x},t)& \rightarrow \Phi _{1}(\mathbf{x},t),
\end{align}%
with similar relationships valid for the molecular field. The Fourier
transforms of the stochastic fields give:%
\begin{align}
\alpha (\mathbf{k},t)& \equiv \mathcal{F}_{\mathbf{x}}[\Psi _{1}(\mathbf{x}%
,t)](\mathbf{k})  \notag \\
& =\frac{1}{(2\pi )^{3/2}}\int d\mathbf{x}\Psi _{1}(\mathbf{x},t)e^{-i%
\mathbf{k\cdot x}} \\
& \rightarrow \frac{1}{(2\pi )^{3/2}}\int d\mathbf{x}\hat{\Psi}_{1}(\mathbf{x%
},t)e^{-i\mathbf{k\cdot x}}=\hat{a}(\mathbf{k},t),  \notag
\end{align}%
\begin{align}
\beta (\mathbf{k},t)& =\mathcal{F}_{\mathbf{x}}[\Phi _{1}(\mathbf{x},t)](%
\mathbf{k})  \notag \\
& =\frac{1}{(2\pi )^{3/2}}\int d\mathbf{x}\Phi _{1}(\mathbf{x},t)e^{-i%
\mathbf{k\cdot x}} \\
& \rightarrow \frac{1}{(2\pi )^{3/2}}\int d\mathbf{x}\hat{\Psi}_{1}^{\dagger
}(\mathbf{x},t)e^{-i\mathbf{k\cdot x}}=\hat{a}^{\dagger }(-\mathbf{k},t). 
\notag
\end{align}

Thus, the Fourier component $\beta (\mathbf{k},t)$ corresponds to $\hat{a}%
^{\dagger }(-\mathbf{k},t)$, rather than to $\hat{a}^{\dagger }(\mathbf{k}%
,t) $ as might have been expected. Therefore, care must be taken in
assigning the operator correspondences in Fourier space.

An alternative choice of the positive-$P$ correspondences between the
operators and the stochastic fields is 
\begin{align}
\hat{\Psi}_{1}(\mathbf{x},t)& \rightarrow \widetilde{\Psi }_{1}^{\ast }(%
\mathbf{x},t), \\
\hat{\Psi}_{1}^{\dagger }(\mathbf{x},t)& \rightarrow \widetilde{\Phi }%
_{1}^{\ast }(\mathbf{x},t).
\end{align}%
These stochastic fields are simply the complex conjugates of the previous
ones: $\widetilde{\Psi }_{1}(\mathbf{x},t)=\Psi _{1}^{\ast }(\mathbf{x},t)$
and $\widetilde{\Phi }_{1}(\mathbf{x},t)=\Phi _{1}^{\ast }(\mathbf{x},t)$.
Their Fourier transforms give:%
\begin{align}
\lbrack \widetilde{\alpha }(\mathbf{k},t)]^{\ast }& =\left\{ \mathcal{F}_{%
\mathbf{x}}[\widetilde{\Psi }_{1}(\mathbf{x},t)](\mathbf{k})\right\} ^{\ast }
\notag \\
& =\frac{1}{(2\pi )^{3/2}}\int d\mathbf{x}\widetilde{\Psi }_{1}^{\ast }(%
\mathbf{x},t)e^{i\mathbf{k\cdot x}} \\
& \rightarrow \frac{1}{(2\pi )^{3/2}}\int d\mathbf{x}\hat{\Psi}_{1}(\mathbf{x%
},t)e^{i\mathbf{k\cdot x}}=\hat{a}(-\mathbf{k},t),  \notag
\end{align}%
\begin{align}
\lbrack \widetilde{\beta }(\mathbf{k},t)]^{\ast }& =\left\{ \mathcal{F}_{%
\mathbf{x}}[\widetilde{\Phi }_{1}(\mathbf{x},t)](\mathbf{k})\right\} ^{\ast }
\notag \\
& =\frac{1}{(2\pi )^{3/2}}\int d\mathbf{x}\widetilde{\Phi }_{1}^{\ast }(%
\mathbf{x},t)e^{i\mathbf{k\cdot x}} \\
& \rightarrow \frac{1}{(2\pi )^{3/2}}\int d\mathbf{x}\hat{\Psi}_{1}^{\dagger
}(\mathbf{x},t)e^{i\mathbf{k\cdot x}}=\hat{a}^{\dagger }(\mathbf{k},t). 
\notag
\end{align}%
Thus, the Fourier component $\hat{a}^{\dagger }(\mathbf{k},t)$ can be
obtained through $[\widetilde{\beta }(\mathbf{k},t)]^{\ast }$ or $\beta (-%
\mathbf{k},t)$, while $\hat{a}(-\mathbf{k},t)$ corresponds to $[\widetilde{%
\alpha }(\mathbf{k},t)]^{\ast }$ and $\alpha (-\mathbf{k},t)$. Using these
correspondences, one can get access to various operator moments involving
the Fourier components $\hat{a}(\pm \mathbf{k},t)$ and $\hat{a}^{\dagger
}(\pm \mathbf{k},t)$. For example, the normal and anomalous densities can be
obtained via%
\begin{eqnarray}
\langle \hat{a}^{\dagger }(\mathbf{k},t)\hat{a}(\mathbf{k},t)\rangle
&=&\langle \lbrack \widetilde{\beta }(\mathbf{k},t)]^{\ast }\alpha (\mathbf{k%
},t)\rangle _{\mathrm{st}}  \notag \\
&=&\langle \beta (-\mathbf{k},t)\alpha (\mathbf{k},t)\rangle _{\mathrm{st}},
\end{eqnarray}%
\begin{align}
\langle \hat{a}(\mathbf{k},t)\hat{a}(-\mathbf{k},t)\rangle & =\langle \alpha
(\mathbf{k},t)[\widetilde{\alpha }(\mathbf{k},t)]^{\ast }\rangle _{\mathrm{st%
}}  \notag \\
& =\langle \alpha (\mathbf{k},t)\alpha (-\mathbf{k},t)\rangle _{\mathrm{st}}.
\end{align}


\begin{thebibliography}{99}
\bibitem{EPR} A. Einstein, B. Podolsky, and N. Rosen, Phys. Rev. \textbf{47}, 777 (1935); N. Bohr, \textit{ibid.} \textbf{48}, 696 (1935).

\bibitem{Kurizki} T. Opatrn\'{y} and G. Kurizki, \prl\textbf{86}, 3180
(2001).

\bibitem{EPR-Bell-comment} We emphasize the distinction to be made between
possible demonstrations of the original EPR paradox and tests of Bell's
inequalities using Bohm's version [D. Bohm, \textit{Quantum Theory},
(Prentice-Hall, Englewoods Cliffs, 1951)] of the EPR correlations for spins
[see, e.g., E. S. Fry, T. S. Walther, and S. Li, \pra \textbf{52}, 4381
(1995)]. Here, we restrict ourselves to references relevant to the original
EPR correlations for continuous variables, and do not attempt to provide an
overview of literature on discrete-variable Bell states or entanglement 
\textit{per se}.

\bibitem{Molecules-bosonic-atoms} R. H. Wynar \textit{et al}., Science 
\textbf{287}, 1016 (2000); E. A. Donley \textit{et al}., Nature (London) 
\textbf{417}, 529 (2002); J. Herbig \textit{et al}., Science \textbf{301},
1510 (2003); K. Xu, T. Muhaiyama, J. R. Abo-Shaeer, J. K. Chin, D. E. Miller, and W. Ketterle, Phys. Rev. Lett. \textbf{91}, 210402
(2003); S. D\"{u}rr, T.Volz, A. Marte, G. Rempe, \textit{ibid.} \textbf{92}, 020406
(2004).

\bibitem{Molecules-fermionic-atoms} C. A. Regal \textit{et al}., Nature
(London) \textbf{424}, 47 (2003); M. Greiner \textit{et al.}, \textit{ibid.} 
\textbf{426}, 537 (2003); 
J. Cubizolles, T. Bourdel, S.J.J.M.F. Kokkelmans, G. V. Shlyapnikov, C. Salomon, Phys. Rev. Lett. \textbf{91}, 240401 (2003); 
K. E. Strecker, G. B. Partridge, R. G. Hulet, \textit{ibid.} \textbf{91}, 080406 (2003); 
M. Zwierlein, C. A. Stan, C. H. Schunk, S. M. F. Raupach, S. Gupta, and Z. Hadzibabbic, \textit{ibid.} \textbf{91}, 250401 (2003).

\bibitem{Durr} S. D\"{u}rr, T. Volz, and G. Rempe, \pra \textbf{70},
031601(R) (2004).

\bibitem{Dissociation-exp-Ketterle} T. Mukaiyama, J. R. Abo-Shaeer, K. Xu, J. K. Chin, and W. Ketterle, Phys. Rev. Lett. \textbf{92}, 180402 (2004).

\bibitem{Greiner} M. Greiner, C. A. Regal, J. T. Stewart, and D. S. Jin, \prl%
\textbf{94}, 110401 (2005).

\bibitem{Moelmer2001} U. V. Poulsen and K. M\o lmer, Phys. Rev. A \textbf{63}, 023604 (2001).

\bibitem{twinbeams} K. V. Kheruntsyan and P. D. Drummond, \pra \textbf{66},
031602(R) (2002); K. V. Kheruntsyan, \pra \textbf{71}, 053609 (2005).

\bibitem{Yurovsky} V. A. Yurovsky and A. Ben-Reuven, Phys. Rev. A \textbf{67}, 043611 (2003).

\bibitem{Jack-Pu} M. W. Jack and H. Pu, \pra \textbf{72}, 063625 (2005).

\bibitem{Fermidiss} K.V. Kheruntsyan, \prl \textbf{96}, 110401 (2006).

\bibitem{KPRL} K. V. Kheruntsyan, M. K. Olsen, and P. D. Drummond, \prl%
\textbf{95}, 150405 (2005).

\bibitem{Ou} Z.Y. Ou, S. F. Pereira, H. J. Kimble, and K. C. Peng, \prl \textbf{68}, 3663 (1992).

\bibitem{eprMDR} M.D. Reid, \pra \textbf{40}, 913 (1989); M.D. Reid and P.D.
Drummond, \prl\textbf{60}, 2731 (1988).

\bibitem{Phillips-4WM} L. Deng \textit{et al.}, Nature (London) \textbf{398}, 218 (1999).

\bibitem{Roberts} D. C. Roberts, T. C. Gasenzer, and K. Burnett,  J. Phys. B \textbf{35}, L113 (2002).

\bibitem{Ketterle-4WM} J. M. Vogels, K. Xu, and W. Ketterle, Phys. Rev.
Lett. \textbf{89}, 020401 (2002); J. M. Vogels, J. K. Chin, and W. Ketterle,
\textit{ibid.} \textbf{90}, 030403 (2003).

\bibitem{Meystre-spin-EPR} H. Pu and P. Meystre, \prl \textbf{85}, 3987
(2000).

\bibitem{Duan-spin-EPR} L-M. Duan, A. Sorensen,  J. I. Cirac, and P. Zoller, \prl \textbf{85}, 3991
(2000).

\bibitem{Soerensen-Duan-Zoller} A. S\o rensen \textit{et al.}, Nature
(London) \textbf{409}, 63 (2001).

\bibitem{Yurovsky-4WM} V. A. Yurovsky, Phys. Rev. A \textbf{65}, 033605
(2002).

\bibitem{Hope} S. A. Haine and J. J. Hope, Phys. Rev. A \textbf{72}, 033601
(2005).

\bibitem{Search-Meystre} C. P. Search and P. Meystre, \textit{Prog.
Opt.} \textbf{47}, 139 (2005).

\bibitem{Chu-4WM-2005} N. Gemelke, E. Sarajlic, Y. Bidel, S. Hong, and S. Chu, Phys. Rev. Lett. \textbf{95}, 170404 (2005).

\bibitem{Ketterle-4WM-2006} G. K. Campbell \textit{et al.}, Phys. Rev. Lett. 
\textbf{96}, 020406 (2006).

\bibitem{Olsen-Davis} M. K. Olsen and M. J. Davis, Phys. Rev. A, 
\textbf{73}, 063618 (2005).

\bibitem{Plata-2005} S. Brouard and J. Plata, Phys. Rev. A \textbf{72},
023620 (2005).

\bibitem{Meystre-diss} T. Miyakawa and P. Meystre, e-print cond-mat/0603469.

\bibitem{Zhao-Astrakharchik} B. Zhao \textit{et al.}, e-print quant-ph/0502011.

\bibitem{Wieman-Julienne-dissociation} S. T. Thompson, E. Hodby, and C. E.
Wieman, Phys. Rev. Lett. \textbf{94}, 020401 (2005); T. K\"{o}hler, E.
Tiesinga, and P. S. Julienne, Phys. Rev. Lett. \textbf{94}, 020402 (2005).

\bibitem{Rempe-Kokkelmans-dissociation} S. D\"{u}rr, T. Volz, N. Syassen, G. Rempe,
E. van Kempen, S. Kokkelmans, B. Verhaar, H. Friedrich, Phys.
Rev. A \textbf{72}, 052707 (2005).

\bibitem{Braaten-2006} E. Braaten and D. Zhang, Phys. Rev. A \textbf{73},
042707 (2006).

\bibitem{Hanna-2006} T. Hanna, K. G\'{o}ral, E. Witkowska, and T. K\"{o}%
hler, Phys. Rev. A \textbf74, 023618 (2006).

\bibitem{Bloch} S. F\"{o}lling \textit{et al.}, Nature (London) \textbf{434}, 481 (2005).

\bibitem{Altman-Lukin} E. Altman, E. Demler, and M. D. Lukin, Phys. Rev. A 
\textbf{70}, 013603 (2004).

\bibitem{Rzazewski} R. Bach and K. Rza\.{z}ewski, Phys. Rev. Lett. \textbf{92%
}, 200401 (2004).

\bibitem{Yasuda-Shimizu} M. Yasuda and F. Shimizu, Phys. Rev. Lett. \textbf{%
77}, 3090 (1996).

\bibitem{Aspect} M. Schellekens \textit{et al.}, Science \textbf{310}, 648
(2005).

\bibitem{Raizen} C.-S. Chuu, F. Schreck, T. P. Meyrath, J. L. Hanssen, G. N. Price, and M. G. Raizen, Phys. Rev. Lett. \textbf{95},
260403 (2005).

\bibitem{Esslinger} A. \"{O}ttl, S. Ritter, M. Kohl, and T. Esslinger, Phys. Rev. Lett. \textbf{95}, 090404 (2005).

\bibitem{HBT} R. Hanbury Brown and R. Q. Twiss, Nature (London) \textbf{177}, 27 (1956).

\bibitem{Duan-full} L. M. Duan, Phys. Rev. Lett. \textbf{96}, 103201 (2006).

\bibitem{Carusotto} Q. Niu, I. Carusotto, and A. B. Kuklov, Phys. Rev. A 
\textbf{73}, 053604 (2006).

\bibitem{Norrie-Ballagh-Gardiner} A. A. Norrie, R. J. Ballagh, and C. W.
Gardiner, Phys. Rev. Lett. \textbf{94}, 040401 (2005); Phys. Rev. A \textbf{73}, 043617 (2006).

\bibitem{Rey-Clark} A. M. Rey, I. I. Satija, and C. W. Clark,
J. Phys. B \textbf{39}, S177 (2006); e-print cond-mat/0604154.

\bibitem{Altman-fermions} L. Mathey, E. Altman, and A. Vishwanath, 
e-print cond-mat/0507108.

\bibitem{Glauber} R. Glauber, Phys. Rev. \textbf{130}, 2529 (1963); see also
M. Naraschewski and R.J. Glauber, Phys. Rev. A \textbf{59}, 4595 (1999).

\bibitem{SavageKheruntsyanSpatial} C. M. Savage and K. V. Kheruntsyan, in
preparation.

\bibitem{DrummondGardiner} P. D. Drummond and C. W. Gardiner, J. Phys. A 
\textbf{13}, 2353 (1980).

\bibitem{DrummondCarter} P. D. Drummond and S. J. Carter, J. Opt. Soc. Am. B 
\textbf{10}, 1565 (1987).

\bibitem{Steel} M. J. Steel, M. K. Olsen, L. I. Plimak, P. D. Drummond, S. M. Tan, M. J. Collett, D. F. Walls, and R. Graham, \pra \textbf{58}, 4824 (1998).

\bibitem{DrummondCorney} P. D. Drummond and J. F. Corney, \pra \textbf{60},
R2661 (1999).

\bibitem{DeuarDrummond} P. Deuar and P. D. Drummond, J. Phys. A 
\textbf{39}, 1163 (2006); \textbf{39}, 2723 (2006).

\bibitem{Gilchrist} A. Gilchrist, C. W. Gardiner, and P. D. Drummond, \pra \textbf{55}, 3014 (1997).

\bibitem{PDKKHH-1998} P. D. Drummond, K. V. Kheruntsyan, and H. He, \prl 
\textbf{81}, 3055 (1998); K.V.Kheruntsyan and P.D.Drummond, Phys. Rev. A 
\textbf{58}, R2676 (1998); P. D. Drummond, K. V. Kheruntsyan, and H. He, J. Optics B: Quant. and Semiclass. Optics \textbf{1}, 387 (1999). 


\bibitem{Superchemistry} D. J. Heinzen, R. Wynar, P. D. Drummond, and K.
V. Kheruntsyan, Phys. Rev. Lett. \textbf{84}, 5029 (2000).

\bibitem{Feshbach-KKPD} P. D. Drummond and K. V. Kheruntsyan, Phys. Rev. A 
\textbf{70}, 033609 (2004).

\bibitem{Timmermans} E. Timmermans, P. Tommasini, R. Cote, M. Hussein, and A. Kerman, Phys. Rev. Lett. \textbf{83}, 2691 (1999); 
Phys. Rep. \textbf{315}, 199 (1999).

\bibitem{JJ-1999} J. Javanainen and M. Mackie, Phys. Rev. A \textbf{59},
R3186 (1999).

\bibitem{Holland} M. Holland, J. Park, and R. Walser, \prl \textbf{86}, 1915
(2001); S. J. J. M. F. Kokkelmans and M. J. Holland, \textit{ibid.} 
\textbf{89}, 180401 (2002).

\bibitem{Stoof-review} R. A. Duine and H. T. C. Stoof, Phys. Rep. \textbf{86}, 115 (2004).

\bibitem{Abrikosov} A. A. Abrikosov, L. P. Gorkov, and I. E. Dzyaloshinski, 
\textit{Methods of Quantum Field Theory in Statistical Physics} (New York,
Dover, 1963).

\bibitem{xmds} G. R. Collecutt and P. D. Drummond, Comp. Phys.
Commun. \textbf{142}, 219 (2001); see also http://www.xmds.org/.

\bibitem{rogue-dissociation-JJ} J. Javanainen and M. Mackie, \prl \textbf{88}, 090403 (2002); 
M. Mackie, K. A. Suominen, and J. Javanainen, \textit{ibid.} \textbf{89}, 180403 (2002).

\bibitem{Comment1} Reducing the statistical noise in the pair correlation in
the regions of $\mathbf{k}$ space with small mode occupancies would require
simulation of a much larger number of stochastic trajectories, and hence a
penalty in the required computational time. The results presented here were
obtained using parallel runs on $20$ dual-processor $3.6$ GHz Dell PowerEdge
servers, taking about $5$ days to collect data for $7500$ trajectory
averages on the computational lattice with $128^{3}$ points and $400$ time steps.

\bibitem{Stochastic-gauges} P. Deuar and P. D. Drummond, Comp. Phys.
Commun. \textbf{142}, 442 (2001); P.  Deuar and P. D. Drummond,
Phys. Rev. A \textbf{66}, 033812 (2002).
\end{thebibliography}
\end{document}